\documentclass{aa}
\usepackage{graphicx}
\usepackage{txfonts}
\usepackage{aalongtable}
\begin{document}
   \title{VLBI observations of the CORALZ sample: young radio sources at low redshift}
   \titlerunning{VLBI observations of the CORALZ sample}

  \author{N. de Vries \inst{1}
    \and
    I. A. G. Snellen \inst{1}
    \and
    R. T. Schilizzi \inst{1,2}
    \and
    K.-H. Mack \inst{3}
    \and
    C. R. Kaiser \inst{4}
  }

  \offprints{vriesn@strw.leidenuniv.nl}

  \institute{Leiden Observatory, Leiden University, P.O. Box 9513, NL-2300 RA, Leiden, The Netherlands\\
                  \email{vriesn@strw.leidenuniv.nl}
    \and
    Square Kilometre Array Program Development Office, Jodrell Bank Centre for Astrophysics, Manchester M13 9PL, UK 
    \and
    Istituto di Radioastronomia - INAF, Via P. Gobetti 101, I-40129 Bologna, Italy
    \and
    School of Physics \& Astronomy, University of Southampton, Southampton SO17 1BJ
  }
 
   \date{Received 2008 October 14; accepted 2008 December 20}

  \abstract
   {Young radio-loud active galactic nuclei form an important tool to investigate the evolution of extragalactic radio sources. To study the early phases of expanding radio sources, we have constructed CORALZ, a sample of 25 compact ($\theta<2''$) radio sources associated with nearby ($z<0.16$) galaxies.}
   {In this paper we determine the morphologies, linear sizes, and put first constraints on the lobe expansion speeds of the sources in the sample.}
   {We observed the radio sources from the CORALZ sample with MERLIN at 1.4 GHz or 1.6 GHz, the EVN at 1.6 GHz, and global VLBI at 1.6 GHz and/or 5.0 GHz.}
   {Radio maps, morphological classifications, and linear sizes are presented for all sources in the CORALZ sample. We have determined a first upper limit to the expansion velocity of one of the sources, which is remarkably low compared to the brighter GPS sources at higher redshifts, indicating a relation between radio luminosity and expansion speed, in agreement with analytical models. 
In addition we present further strong evidence that the spectral turnovers in GPS and CSS sources are caused by synchrotron self-absorption (SSA): the CORALZ sources are significantly offset from the well-known correlation between spectral peak frequency and angular size, but this correlation is recovered after correcting for the flux-density dependence, as predicted by SSA theory.}
   {}


   \maketitle
%

\section{Introduction} \label{COR_int}

Young radio-loud active galactic nuclei (AGN) are ideal objects to study the trigger of AGN activity, and the early evolution of classical double radio sources. Young radio sources can be recognised by their small angular (physical) sizes, their inverted spectra at low frequencies, and their often compact symmetric morphologies, with jet or lobe-like structures on either side of a central core. Depending on the observed characteristics, they are called Compact Steep Spectrum (CSS) sources (Peacock \& Wall \cite{Peacock82}; Fanti et al. \cite{Fanti90}; O'Dea \cite{O'Dea98}), Gigahertz Peaked Spectrum (GPS) sources (O'Dea \cite{O'Dea98}), and/or Compact Symmetric Objects (CSO, Wilkinson et al. \cite{Wilkinson94}; Conway \cite{Conway02}).

Samples of CSS, GPS, and CSO have been used in a statistical way to constrain the early evolution of radio sources. 
Relative number statistics indicated that at a young age, radio sources must have significantly higher radio luminosities than at a later stage (O'Dea \& Baum \cite{O'DeaBaum96}; Readhead et al. \cite{Readhead96}; Fanti \& Fanti \cite{Fanti03}). This decrease in radio luminosity may be preceded by a period of luminosity increase (Snellen et al. \cite{Snellen00b}; \cite{Snellen03}).

Very interesting results come from multi-epoch Very Long Baseline Interferometry (VLBI) observations of individual GPS and CSO. They show, for a handful of powerful objects the proper motions of the lobes, from which their dynamical ages can be estimated (Owsianik \& Conway \cite{Owsianik98}; Polatidis \& Conway \cite{Polatidis03} and references therein). Separation velocities between the opposite extremities of the sources of up to $0.4 h^{-1} c$ have been measured, corresponding to dynamical ages in the range of a few hundred to a few thousand years.

\subsection*{The CORALZ sample: The most nearby young radio sources}

Snellen et al. (\cite{CORALZ}) have selected a sample of young radio sources, CORALZ (COmpact RAdio sources at Low Redshift), with the aim of obtaining an unbiased view of young radio-loud AGN in the nearby universe. The sources in the CORALZ sample are significantly closer than the archetypal CSO and GPS sources available in the literature. Proper motions should be easier to detect in these nearby sources, since similar intrinsic expansion velocities would result in significantly higher angular motions compared to previous studies. Moreover, in general these radio sources are less powerful than the archetypal CSO and GPS sources, which will allow luminosity-dependent comparison studies.

The CORALZ programme was also initiated to investigate possible selection biases in GPS and CSO samples, since CORALZ is thought to be much less biased, due to its relatively clean selection criteria. The sample has been selected on flux density ($S_{\rm{1.4\ GHz}}>$ 100 mJy) and angular size ($\theta<2''$). Using the Very Large Array (VLA) Faint Images of the Radio Sky at Twenty-centimeters (FIRST) survey (White et al. \cite{White97}), the optical Automated Plate Measuring machine (APM) catalogue of the first Palomar Observatory Sky Survey (POSS-I) (McMahon \& Irwin \cite{McMahon91}), and follow-up observations, all radio sources identified with bright galaxies were selected (red magnitude of $e < 16.5$~mag \textit{or} blue magnitude of $o < 19.5$~mag).
Originally, four of these 28 sources where excluded from the CORALZ sample, since the offset between the radio position and its optical counterpart was suspiciously large ($\Delta_{\rm{pos}}>2''$), suggesting that the radio source had been erroneously identified with a random foreground galaxy. In the mean time, based on the Sloan Digital Sky Survey (SDSS) Data Release 6 (Adelman-McCarthy et al. \cite{Adelman-McCarthy08}), one of these identifications actually turned out to be correct (\object{CORALZ J115000+552821} and \object{SDSS J115000.08+552821.4}, with an offset of $0.5''$), raising the number of radio sources with reliable optical identifications to 25. These 25 compact radio sources will be referred to as \textsl{the CORALZ sample}. The three sources that still show anomalously large offsets between the radio and optical position, are referred to as \textsl{additional sources}.
Although the CORALZ sample is statistically complete down to the flux density limit, the limit corresponds to different redshifts, since the host galaxies are selected on optical magnitude and have a range in absolute magnitude. However, simulations show that the sample is 95\% statistically complete in the redshift range $0.005<z<0.16$ (Snellen et al. \cite{CORALZ}), which, including the new SDSS identification, now contains 18 sources. In the rest of this paper we will refer to these 18 sources as \textsl{the CORALZ core sample}. A detailed description of the selection process and additional radio observations can be found in Snellen et al. (\cite{CORALZ}).
Nearly all (17/18) of the sources in the CORALZ core sample are classified as GPS and CSS sources, confirming that these are likely to be young radio sources. This also confirms that the spectral shape of a radio source is a good criterion to select very young radio (GPS) sources in general.

In this paper we present VLBI and MERLIN observations of the CORALZ sample. In Sect. \ref{COR_obs} the observations and data reduction are described; the data analysis and modeling is described in Sect. \ref{COR_ana}. The results are presented and discussed in Sect. \ref{COR_res}, and we conclude with Sect. \ref{COR_con}. Throughout the paper we adopt the cosmological parameters as found by WMAP5 (Komatsu et al. \cite{Komatsu08}; $H_0 = 70.1\ \rm km~s^{-1}~Mpc^{-1}$, $\Omega_\Lambda = 0.721$, $\Omega_{\rm{m}} = 0.279$).

\section{Observations and data reduction} \label{COR_obs}

We observed radio sources from the CORALZ sample using a range of arrays at several frequencies; MERLIN, the Multi-Element Radio Linked Interferometer Network, at 1.4 GHz or 1.6 GHz, the European VLBI Network (EVN) at 1.6 GHz, and global VLBI at 1.6 GHz and/or 5.0 GHz. The combination of telescope array and observing frequency were chosen to have a wide range in angular resolution, required to match the expected wide range in angular size of the sources in the sample. An initial guess of the angular sizes of the individual sources was obtained from the position and strength of the radio spectral peak. In this way the sample was divided into three sub-samples of expected `large', `intermediate', and `small' size sources. Details of the different observing runs are given below.

\subsection*{MERLIN observations}
The subsample of `large' radio sources (seven sources, including four from the CORALZ core sample) was observed with MERLIN using the following six telescopes: Defford, Cambridge, Knockin, Darnhall, Jodrell Bank Mk2, and Pickmere. The observing times ranged from 9 to 17 hours, alternating between the science targets (3 min.) and the radio sources used as their phase-references (2 min.), resulting in on-source integration times ranging from 5 to 10 hours. The largest projected baselines were typically 200 km, resulting in a resolution of about 150 mas using uniform weighting. These observations were carried out between November 9, 2000, and January 2, 2001, at either 1.4 GHz or 1.6 GHz.

\subsection*{EVN observations at 1.6 GHz}
The twelve `intermediate' size CORALZ sources (including ten sources from the CORALZ core sample) were observed with the EVN at 1.6 GHz using the Effelsberg, Jodrell Bank, Medicina, Noto, Torun, Westerbork, and Onsala antennas between February 14, 2000, and May 28, 2001, except for two sources, \object{CORALZ J134158+541524} and \object{CORALZ J150805+342323}, which were observed on February 19, 2002, using the MERLIN antenna at Cambridge instead of Noto.
Each source was typically observed for $10 \times 13$ minutes, spread in time to obtain optimal {\it UV} coverage. The largest projected baselines were typically 1400 km, resulting in a resolution of about 20 mas using uniform weighting.
J0650+6001 and J1740+5211 were observed as primary calibrators. The data correlation was performed at Socorro on the VLBA correlator.
The Astronomical Image Processing System (AIPS) has been used for editing, a-priori calibration, fringe-fitting, self-calibration and imaging of the data. For fringe-fitting, the AIPS task FRING was used with a point source model, a solution interval of 6 minutes, and a standard signal-to-noise ratio cutoff of 5. Several iterations of phase (or amplitude \& phase) self-calibration and imaging were performed using the AIPS tasks CALIB and IMAGR, until the image quality converged, with negligible negative structure and a low noise level.

\subsection*{Global VLBI observations (2000) at 5.0 GHz}
The eight `small' CORALZ sources (including three sources from the CORALZ core sample) were observed on March 2, 2000, with an 18 station array at a frequency of 5.0 GHz, using the EVN antennas Effelsberg, Jodrell Bank, Medicina, Noto, Torun, Westerbork, Onsala, and Shanghai and the Very Long Baseline Array (VLBA) antennas St. Croix, Hancock, North Liberty, Fort Davis, Los Alamos, Pie Town, Kitt Peak, Owens Valley, Brewster, and Mauna Kea.
Each source was typically observed for $3 \times 11$ minutes, spread in time to obtain optimal {\it UV} coverage. The largest projected baselines were typically 9000 km, resulting in a resolution of about 2 mas using uniform weighting.
J0650+6001 and J1740+5211 were observed as primary calibrators. The data correlation was performed at Socorro on the VLBA correlator. The data reduction was carried out the same way as the EVN observations.

\subsection*{Global VLBI observations (2004) at 1.6 GHz and 5.0 GHz}
The fourteen most compact sources from the CORALZ core sample were observed for a second time on May 24, 2004, and June 5, 2004, at 5.0 GHz and/or 1.6 GHz, respectively. A similar observing strategy was used as the previous global VLBI observations, with observing times of $3 \times 11$ minutes. The largest projected baselines were typically 9000 km, resulting in a resolution of about 2 mas at 5.0 GHz, and 5 mas at 1.6 GHz, using uniform weighting.
Both experiments were conducted with global VLBI, using the EVN antennas Effelsberg, Jodrell Bank, Medicina, Noto, Torun, Westerbork, and Onsala and the VLBA antennas St. Croix, Hancock, North Liberty, Fort Davis, Los Alamos, Pie Town, Kitt Peak, Owens Valley, and Brewster. J1740+5211 and 4C39.25 were observed as primary calibrators. The data correlation was performed at Joint Institute for VLBI in Europe (JIVE). The data reduction was carried out the same way as the EVN observations.

All details of the observations are given in the appendix in Table \ref{TCOR_1}; in Col. 1 IAU name of the source, in Col. 2 the figure in which the map is presented, in Cols. 3-7 the observing date, frequency, bandwidth, and array, in Col. 8 the peak flux density, in Col. 9 the r.m.s. noise level, and in Cols. 10-12 the major and minor axis and the position angle of the restoring beam.

\section{Data analysis} \label{COR_ana}

\subsection{Source structure and angular sizes} \label{COR_anaSS}
The observed radio source structures were characterised using the AIPS task JMFIT, which fits (by least-squares) up to four Gaussian components to an image subsection. A first guess of the input parameters (position, flux density, and shape of the components) was made using the INPFIT procedure. The results for all observations of all sources are given in Table \ref{TCOR_3}, with in Col. 1 the source IAU name, in Col. 2 the figure in which the corresponding map is presented, in Col. 3 the observation date, in Col. 4 the observing frequency, in Cols. 5-6 the relative position of the component, in Cols. 7-9 the flux density, deconvolved size and position angle of the component. 
The angular sizes of the radio sources were determined by measuring the distance between the outer edges of the two outermost components in the deconvolved image. In the case of unresolved components, the positions of the components were used. For unresolved or barely resolved sources the deconvolved major axis as returned by JMFIT was used as an estimate of the angular size of the radio source.

\subsection{Morphological classification} \label{COR_anaMor}
We also made an attempt to morphologically classify the radio sources by eye, using the source structures at different frequencies. We identify five classes:
\begin{enumerate}
\item Unresolved (U). The source is unresolved or barely resolved.
\item Compact Double (CD). The source exhibits two components with a similar spectral index and/or flux density.
\item Compact Symmetric Object (CSO). A source with a range of components, with the most flat spectral component \textsl{not} located at one of the extremities of the source.
\item Core-Jet (CJ). A source with two or more components, significantly different in flux density and/or spectral index, with the most flat spectral component located at one of the extremities of the source.
\item Complex (CX). A source with a complex morphology, not falling in one of the above categories.
\end{enumerate}
Each identification is followed by a `?' if the classification is particularly uncertain. The results are shown in Table \ref{TCOR_2}, and discussed in Sect. \ref{COR_res_morph}.

\subsection{Expansion velocity}
For three sources we have obtained two epochs of VLBI observations (Fig. \ref{FCOR_2}), observed with the same telescope array at the same frequency, for which we could potentially estimate expansion velocities. One of these sources (\object{CORALZ J073934+495438}) is practically unresolved, and a second source (\object{CORALZ J131739+411545}) shows a complex morphology, making the analysis of expansion velocities for these sources impossible at this stage. The third source (\object{CORALZ J083139+460800}) on the other hand exhibits a compact double morphology with two strong, unresolved components separated by 4.4 mas. The relative positions of the two components were determined using the AIPS tasks JMFIT as well as MAXFIT, which fits a quadratic function to a selected region of the map to determine the position of an extremum. It is always difficult to obtain error estimates for component positions from VLBI observations, which are particularly important in case of slow expansion speeds and upper limits. When three or more epochs of observations are available, positional errors are often derived from the scatter around a linear fit to the component motion (Owsianik \& Conway \cite{Owsianik98}), although in this case no independent goodness-of-fit can be determined. Since we only have two epochs of observations, we use the following formula from Fomalont (\cite{Fomalont99}) to derive estimates of the error in the component position:
\begin{equation} \label{COR_eqPosErr}
  \sigma_r = \frac{\sigma_{\rm{rms}}\ d}{2\ I_{\rm{peak}}}
\end{equation}
where $\sigma_{\rm{rms}}$ is the post-fit r.m.s. error of the map, and $d$ and $I_{\rm{peak}}$ are the size and the peak intensity of the component. We have also used Monte Carlo simulations to check our error estimates, by repeatedly fitting the peak position of a gaussian with random noise added. The error estimates from the approximating formula from Fomalont (\cite{Fomalont99}) and from the Monte Carlo simulations were very similar. We do note, however, that these error estimates only take into account the errors in the analysis of the radio maps, not the possible biases introduced during the synthesis of these maps.

\section{Results and discussion} \label{COR_res}

\begin{table*}
\centering
\caption{Source characteristics of the CORALZ sample.\label{TCOR_2}}
\begin{tabular}{ccrcrrrc}
\hline
IAU Name & $z$ & $S_{\rm{1.4 GHz}}$        & $L_{\rm{5.0 GHz}}$ & 
\multicolumn{1}{c}{$\nu_{\rm{peak}}$} & \multicolumn{1}{c}{$\theta$} & LLS                      & Morph. \\
         &     & \multicolumn{1}{c}{(mJy)} & (W Hz$^{-1}$)      & 
(MHz)               & (mas)                & \multicolumn{1}{c}{(pc)} &        \\
\hline
\multicolumn{8}{l}{The CORALZ core sample} \\
\object{J073328+560541} & 0.104 & 394 & 24.68 & 460    &   47 &   90 & CSO \\
\object{J073934+495438} & 0.054 & 107 & 23.63 & 950    &   2* &   2* & U \\
\object{J083139+460800} & 0.127 & 131 & 24.62 & 2200   &    9 &   20 & CD \\
\object{J083637+440109} & 0.054 & 139 & 23.66 & $<150$ & 1600 & 1700 & CSO? \\
\object{J090615+463618} & 0.085 & 314 & 24.49 & 680    &   31 &   49 & CSO \\
\object{J102618+454229} & 0.153 & 105 & 24.55 & 180    &   17 &   45 & CSO \\
\object{J103719+433515} & 0.023 & 129 & 22.96 & $<150$ &   19 &    9 & CSO \\
\object{J115000+552821} & 0.139 & 143 & 24.57 & $<230$ &   41 &  100 & U \\
\object{J120902+411559} & 0.095 & 147 & 24.26 & 370    &   20 &   35 & CSO \\
\object{J131739+411545} & 0.066 & 249 & 24.37 & 2300   &    4 &    5 & CX \\
\object{J140051+521606} & 0.116 & 174 & 24.36 & $<150$ & 150* & 320* & U \\
\object{J140942+360416} & 0.148 & 143 & 24.45 & 330    &   27 &   70 & CJ? \\
\object{J143521+505122} & 0.099 & 141 & 24.20 & $<150$ & 150* & 270* & U \\
\object{J150805+342323} & 0.045 & 130 & 23.35 & $<230$ &  170 &  150 & CD? \\
\object{J160246+524358} & 0.106 & 576 & 24.75 & $<150$ &  180 &  350 & CSO \\
\object{J161148+404020} & 0.152 & 553 & 25.03 & $<150$ & 1300 & 3400 & CX \\
\object{J170330+454047} & 0.060 & 119 & 23.54 & $<150$ &      &      &  \\
\object{J171854+544148} & 0.147 & 329 & 24.86 & 480    &   68 &  175 & CSO? \\
\multicolumn{8}{l}{Other nearby sources in the CORALZ sample} \\
\object{J093609+331308} & 0.076 &  55 & 23.84 & 2200   & 1.5* &   2* & U \\
\object{J101636+563926} & 0.232 & 108 & 24.91 & $<150$ &  240 &  890 & CD? \\
\object{J105731+405646} & 0.008 &  47 & 21.59 & 1250   & 0.5* & 0.1* & U \\
\object{J115727+431806} & 0.229 & 256 & 25.25 & $<150$ &  630 & 2300 & CJ? \\
\object{J132513+395552} & 0.074 &  56 & 23.69 & 1900   &   10 &   14 & CSO? \\
\object{J134035+444817} & 0.065 &  82 & 23.89 & 2300   &  3.3 &  4.1 & CJ? \\
\object{J155927+533054} & 0.178 & 182 & 24.67 & $<150$ & 1500 & 4500 & CSO? \\
\hline
\multicolumn{8}{l}{Additional sources} \\
\object{J071509+452555} &  &  74 & & 3800   &  1.3 &  & CD? \\
\object{J080454+433537} &  & 360 & & 1500   &   24 &  & CD? \\
\object{J134158+541524} &  & 125 & & $<150$ &  530 &  & CD? \\
\hline
\end{tabular}
\end{table*}

\subsection{Source structure}
Contour maps of the CORALZ sample at 1.4, 1.6, and/or 5.0 GHz, using the EVN, global VLBI, and MERLIN arrays, are shown in the appendix. Fig. \ref{FCOR_1} shows contour maps of those eight sources from the CORALZ core sample, observed with the EVN at 1.6 GHz in 2000/2001 (left column), and with global VLBI in 2004 at 1.6 GHz and 5.0 GHz (middle and right column, respectively). Fig. \ref{FCOR_2} shows contour maps of the three most compact sources in the CORALZ core sample, obtained with global VLBI at 5.0 GHz in 2000 (top row), and 2004 (bottom row). Fig. \ref{FCOR_3} shows radio maps of the six more extended sources in the CORALZ core sample, taken with MERLIN in 2000 at 1.4 GHz or 1.6 GHz, or with the EVN at 1.6 GHz. Fig. \ref{FCOR_4} shows contour maps of other sources from the CORALZ sample, that are not part of the core sample, obtained with global VLBI at 5.0 GHz or with MERLIN at 1.4 GHz or 1.6 GHz. These objects are nearby GPS/CSS radio sources, but are not included in the CORALZ core sample, because they either have slightly too low flux densities ($S_{\rm 1.4\ GHz}<100$ mJy), or are too distant ($z>0.16$). Fig. \ref{FCOR_5} shows contour maps of the additional sources, which have been excluded from the CORALZ sample because the optical and radio positions are significantly offset, indicating that the radio sources have been erroneously identified with a random foreground galaxy. For three sources (\object{CORALZ J073328+560541}, \object{CORALZ J103719+433515}, and \object{CORALZ J160246+524358}), the S/N at 5.0 GHz on the longer baselines was too low for fringe-fitting, and we used the EVN antennas only. One source (\object{CORALZ J140942+360416}) was too weak at 5.0 GHz for self-calibration, and we were only able to produce a map at 1.6 GHz.

From these observations we have derived the largest angular size, $\theta$, of each source, from which the (projected) largest linear size (LLS) can be calculated. Six sources (\object{CORALZ J073934+495438}, \object{CORALZ J115000+552821}, \object{CORALZ J140051+521606}, \object{CORALZ J143521+505122}, \object{CORALZ J093609+331308}, and \object{CORALZ J105731+405646}) were (almost) unresolved by our observations, and we have estimated $\theta$ using the deconvolved major axis. Sizes determined this way are less certain and are therefore indicated with a star. The results are given in Table \ref{TCOR_2}, with in Col. 1 the source IAU name, 
in Cols. 2-5 the redshift, the flux density at 1.4 GHz, the luminosity density at 5.0 GHz, and the peak frequency (Snellen et al. \cite{CORALZ}),
 in Col. 6 the largest angular size $\theta$, in Col. 7 the projected largest linear size, in Col. 8 a possible morphological classification (see Sect. \ref{COR_anaMor} for a description of the different classes).

\subsection{Radio morphology} \label{COR_res_morph}
For 13 out of the 18 sources from the CORALZ core sample, a morphological classification was possible. The majority of the classified CORALZ sources show a CSO or CD morphology (10/13, 77\%). Only one source has been tentatively classified as a core-jet source. These results are in agreement with Stanghellini et al. (\cite{Stanghellini97}) and Snellen et al. (\cite{Snellen00a}), who find that GPS sources classified as core-jet are usually identified with quasars, whereas GPS sources identified with galaxies tend to exhibit CSO morphologies.

\subsection{Spectral turnover and angular size}
\begin{figure*}[th]
\centering
\hbox{
 \includegraphics[width=8.5cm]{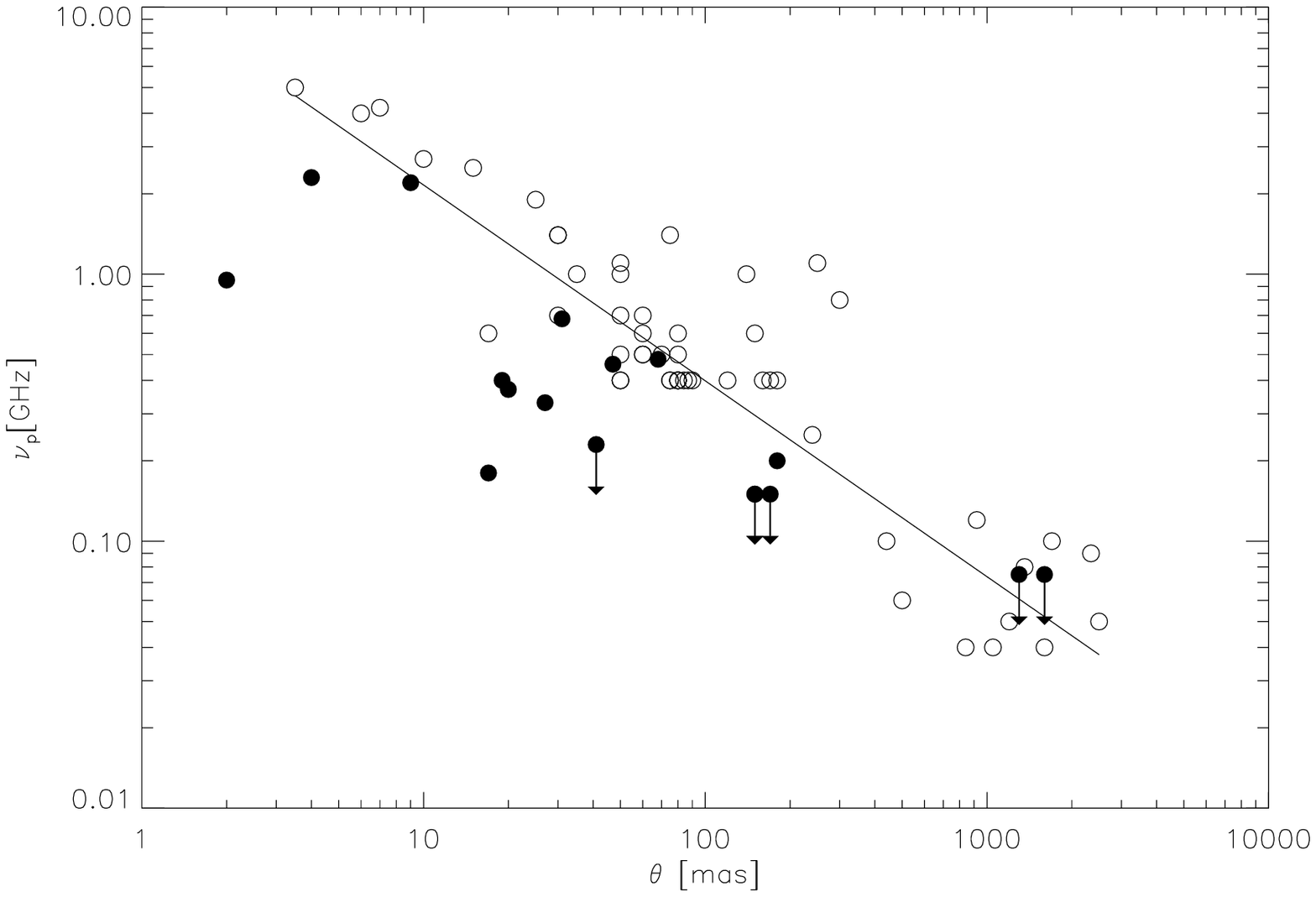}
 \includegraphics[width=8.5cm]{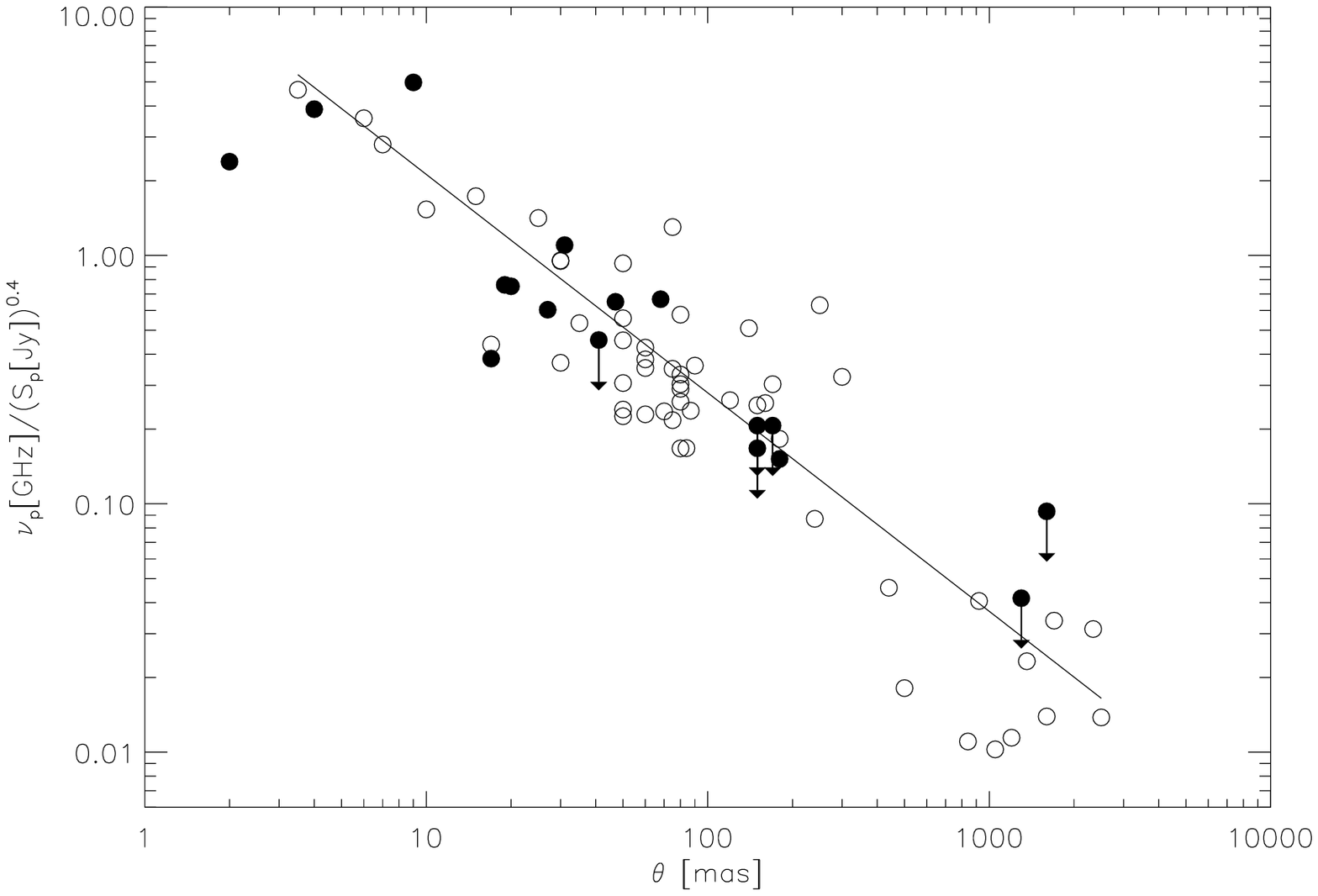}
}
\caption{({\it left}) Largest angular sizes and peak frequencies of the CORALZ core sample (filled cicles) and the bright PHJ (Snellen et al. 2002), Stanghellini (1998), and Fanti (1990) samples (open circles). Circles with arrows denote upper limits. The line is a linear least-squares fit (in log-log space) to the bright literature samples. ({\it right}) To correct for the flux density-dependence predicted by SSA, we divide the peak frequency by $S_m^{2/5}$, and show that the linear sizes and peak frequencies of the CORALZ sources are now consistent with the linear least-square fit to the bright samples.}
\label{FCOR_LLS1}
\end{figure*}

Following Snellen et al. (\cite{Snellen00b}) we searched for correlations between the angular sizes and spectral turnovers of the sources in the CORALZ core sample, and compared the results with those presented in the literature.
Early measurements of the angular sizes of GPS and CSS sources using VLBI strongly suggested that their spectral turnovers are caused by synchrotron self-absorption (SSA, Jones, O'Dell \& Stein \cite{Jones74}; Hodges, Mutel \& Phillips \cite{Hodges84}; Mutel, Hodges \& Phillips \cite{Mutel85}).
This was further advocated by Fanti et al. (\cite{Fanti90}), who showed that the peak frequency and the largest angular size of CSS radio sources are strongly anticorrelated, as expected for SSA. For a homogeneous synchrotron self-absorbed radio source, the peak frequency, $\nu_{\rm{p}}$, is given approximately by (Kellermann \& Pauliny-Toth \cite{Kellermann81}):
\begin{equation} \label{COR_eqSSA}
  \nu_{\rm{p}} \sim 8\ B^{1/5}\ S_{\rm{p}}^{2/5}\ \theta^{-4/5}\ (1+z)^{1/5}\ \mathrm{GHz},
\end{equation}
where $B$ is the magnetic field in gauss, $S_{\rm{p}}$ the flux density at the spectral peak in Jy, $\theta$ the largest angular size in mas, and $z$ the redshift of the source. In contrast, Bicknell et al. (\cite{Bicknell97}) argue that models where the radio source is surrounded by ionized gas can also reproduce the observed relation, and the turnover could also be caused by free-free absorption (FFA). 

Now that we have measured sizes and peak frequencies of the CORALZ core sample, at significantly lower redshifts and radio luminosities than the literature samples, it is interesting to see whether weak radio sources obey the same anticorrelation between $\theta$ and $\nu_{\rm{p}}$.
We have collected angular sizes and peak frequencies of several flux density-selected samples from the literature for comparison. These included the Parkes half-Jansky (PHJ) sample of GPS galaxies (Snellen et al. \cite{Snellen02}, with angular sizes from Liu et al. \cite{Liu07}), the GPS galaxies from the Stanghellini (\cite{Stanghellini98}) sample, and the CSS galaxies from the Fanti (\cite{Fanti90}) sample. The angular sizes for the sources in the latter two samples come from O'Dea (\cite{O'Dea98}).
 
Fig. \ref{FCOR_LLS1} ({\it left}) shows that the peak frequency and linear size of the CORALZ core sample (filled circles) are also anticorrelated, but they fall significantly below the relation for the brighter and more distant literature samples. However, this is actually expected from SSA theory. As can be seen in equation (\ref{COR_eqSSA}), the peak frequency of a synchrotron self-absorbed radio source scales with the flux density as: $\nu_{\rm{p}} \propto S_{\rm{p}}^{2/5}$, and therefore the low flux density CORALZ sample should have smaller angular sizes for similar peak frequencies. In Fig. \ref{FCOR_LLS1} ({\it right}) we have corrected for this effect, and show that the linear sizes and peak frequencies of the CORALZ core sample are now consistent with the linear fit to the bright literature samples, perfectly in line with synchrotron self-absorption theory.

\begin{figure}[ht]
\centering
 \includegraphics[width=8.5cm]{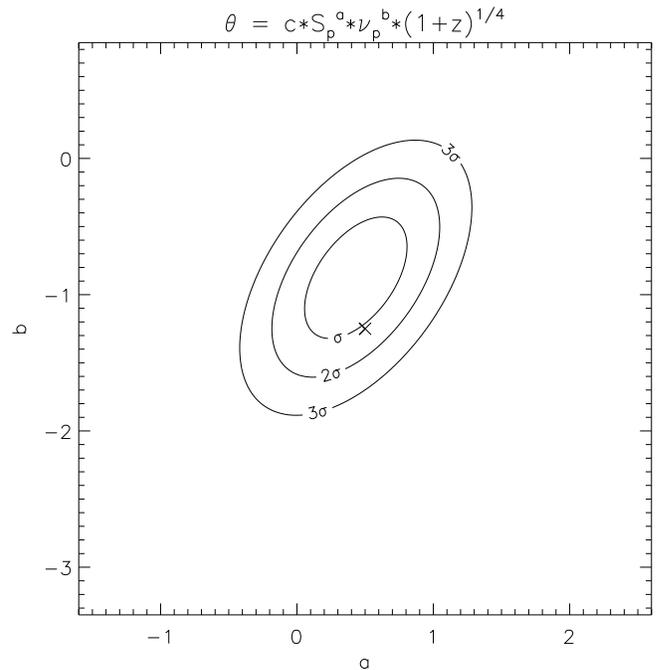}
\caption{Contour map of the goodness-of-fit parameter $\Delta\chi^2$ for the combined sample of young radio sources (the CORALZ core sample, PHJ (Snellen et al. 2002), Stanghellini (1998), Fanti (1990)). The contours are drawn at the 68.3\%, 95.4\% and 99.7\% levels of a $\Delta\chi^2$-distribution with two degrees of freedom. The location of the parameters expected for SSA (equation (\ref{COR_eqSSA})) is indicated by the cross.}
\label{FCOR_LLS2}
\end{figure}

In Fig. \ref{FCOR_LLS1} ({\it right}) we assume that GPS and CSS radio sources obey equation (\ref{COR_eqSSA}), although this formula in principle only applies to homogeneous synchrotron self-absorbed sources. 
From VLBI observations it is clear that GPS/CSS sources are not homogeneous, so the fact that the correlation between peak frequency and overall angular size is so tight, suggests that the ratio of component to overall size does not change significantly over the lifetime of GPS/CSS radio sources. This fixed ratio of component to overall size is in agreement with VLBI observations and can be explained as self-similar evolution (e.g. Snellen et al. \cite{Snellen00b}).
To determine how the data constrain the correlation between turnover frequency, peak flux density and angular size to be in agreement with SSA and self-similar evolution, we fit the data using the fitting function:
\begin{equation} \label{COR_eqSSA2}
  \theta = c\ S_{\rm{p}}^{a}\ \nu_{\rm{p}}^{b}\ (1+z)^{1/4},
\end{equation}
where $a$, $b$, and $c$ are parameters that can be varied in order to best fit the data. For $a=1/2$, $b=-5/4$, and $c=8^{5/4}B^{1/4}$, this expression reduces to equation (\ref{COR_eqSSA}). In Fig. \ref{FCOR_LLS2} we show a contour map of the goodness-of-fit parameter $\Delta\chi^2$, resulting from fitting the data to equation (\ref{COR_eqSSA2}) with a range of fixed values for $a$ and $b$. We again used the CORALZ core sample, and the PHJ (Snellen et al. 2002), Stanghellini (1998), and Fanti (1990) samples. Note that the five larger sources from the CORALZ core sample, which have upper limits to the turnover frequency (see Fig. \ref{FCOR_LLS1}), are not included in this analysis. When turnover frequencies and peak flux densities for these become available, we expect to significantly improve the constraints on the parameters in equation (\ref{COR_eqSSA2}). The contours are drawn at the 68.3\%, 95.4\% and 99.7\% levels of a $\Delta\chi^2$-distribution with two degrees of freedom. The values of the parameters $a$ and $b$ expected for a homogeneous synchrotron self-absorbed source (equation (\ref{COR_eqSSA})) are indicated by the cross. From Fig. \ref{FCOR_LLS2} we conclude that the data are in remarkably good agreement with the homogeneous SSA source model, with $a=0.45 \pm 0.25$ and $b=-0.9 \pm 0.3$. This confirms that the spectral turnovers in GPS and CSS spectra are generally caused by SSA, and also favours self-similar evolution models of young radio sources.

\begin{figure}[ht]
\centering
 \includegraphics[width=8.5cm]{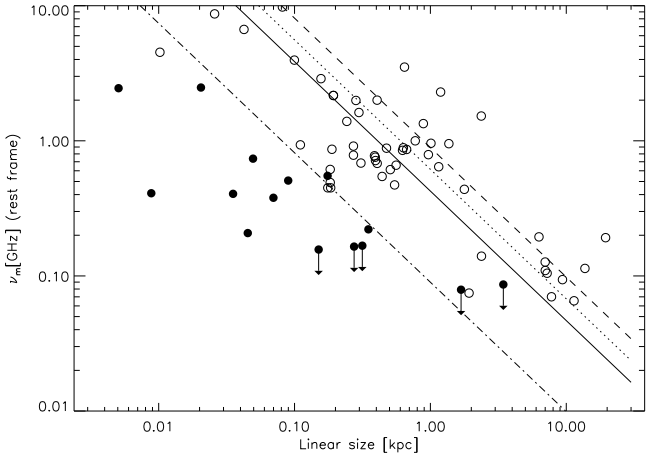}
\caption{Rest-frame turnover frequency against linear size for the CORALZ core sample (filled cicles) and the PHJ (Snellen et al. 2002), Stanghellini (1998), and Fanti (1990) samples (open circles). The lines are models from Bicknell et al. (1997) for jet energy fluxes of $10^{46}$, $10^{45.5}$, $10^{45}$, and $10^{42.7} {\rm erg\ s}^{-1}$ (dashed, dotted, solid, and dot-dashed line, respectively), density power-law slope of $\beta=2$, and number density at 1 kpc of $10\ {\rm cm}^{-3}$.}
\label{FCOR_LLS3}
\end{figure}

Can FFA also explain the relations found? Bicknell et al. (\cite{Bicknell97}) present radio source evolution models, where the AGN ionizes the gas surrounding the radio source by shocks and photoionization. These models predict an anticorrelation between the rest-frame peak frequency and the largest linear size of GPS and CSS radio sources, consistent with the data available at that time (the samples of Fanti et al. (\cite{Fanti90}) and Stanghellini et al. (\cite{Stanghellini98})). To see whether the sizes and peak frequencies of the CORALZ core sample are also in agreement with this model, we have reproduced Fig. 7 ({\it top right}) from Bicknell et al. (\cite{Bicknell97}) in Fig. \ref{FCOR_LLS3}, and added the data from the CORALZ core sample. Sources from the CORALZ core sample are represented by filled cicles and sources from the PHJ (Snellen et al. 2002), Stanghellini (1998), and Fanti (1990) samples by open circles. The lines are predictions of the radio source models from Bicknell et al. (1997) for jet energy fluxes of $10^{45}$, $10^{45.5}$, and $10^{46} {\rm erg\ s}^{-1}$ (solid, dotted, and dashed line, respectively), density power-law slope of $\beta=2$, and number density at 1 kpc of $10\ {\rm cm}^{-3}$. The fact that the lines fit the literature samples fairly well is no surprise, since the model is designed to do so.

The median radio luminosity of the CORALZ core sample is a factor of 630 lower than that of the bright literature samples, meaning that the jet energy fluxes should also be a factor of 630 lower, assuming a constant fraction of the jet energy radiated as synchrotron emission (as expected from analytical models, e.g. see appendix Sect. \ref{COR_model}), resulting in a jet energy flux of $\sim 10^{42.7} {\rm erg\ s}^{-1}$. This is indicated by the dot-dashed line in Fig. \ref{FCOR_LLS3}. All-but-one of the CORALZ core members with measured turnover frequency fall below this line, indicating that the model is not consistent with these data, or that, for some reason, the ratio of jet energy flux to radio luminosity is significantly lower for the CORALZ core sample. Of course, it will be possible to add extra parameters to the Bicknell model, such that it fits the peak frequency and angular size data of GPS and CSS sources. However, since the SSA self-similar evolution scenario is so straightforward and fits the data well, we believe it is scientifically more elegant.

\subsection{Expansion velocity and radio power} \label{COR_VexpL}
\begin{figure}[ht]
\centering
 \includegraphics[width=8.5cm]{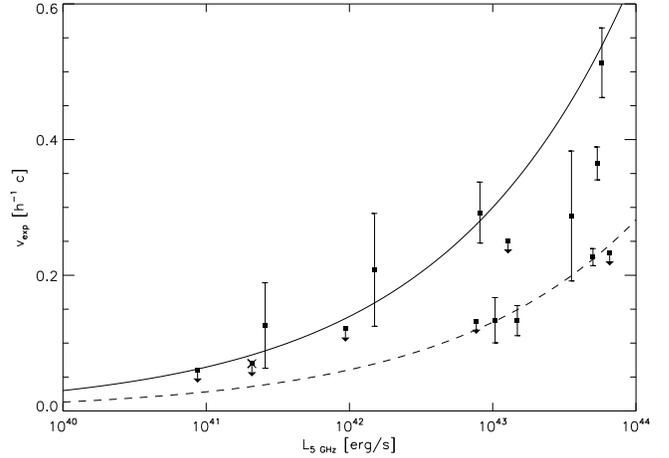}
\caption{Expansion velocities of Compact Symmetric Objects in units of $h^{-1}\ c$, and radio luminosities at 5 GHz in $\rm erg\ s^{-1}$. The upper limit indicated with the cross is our 3$\sigma$ upper limit to the expansion velocity of \object{CORALZ J083139+460800}, the other arrows indicate 1$\sigma$ upper limits from previous studies. The solid line shows the expected relation between expansion velocity and radio luminosity, derived from an analytic radio source evolution model described in the appendix in Sect. \ref{COR_model}, with arbitrary scaling. The dashed line represents this relation scaled down by a factor of $\sqrt{1-0.9^2} \simeq 0.44$, such that 90\% of a sample of randomly oriented sources that intrinsically would follow the solid line, should be located above the dashed line.}
\label{FCOR_vL}
\end{figure}

Three sources have been observed twice (March 2000 and May 2004) with the same VLBI array at the same frequency, allowing investigation of possible structural changes, but only one source, \object{CORALZ J083139+460800}, is suitable for determination of the expansion velocity (see above). However, we found no change in the position ($x$ or $y$) to within 18 $\mu$as, which corresponds to a 3$\sigma$ upper limit to the projected expansion velocity of $v_{\rm exp} <$ 0.07 $h^{-1} c =$ 0.10 $c$. We noted that this upper limit is lower than the expansion velocities previously found for bright archetypal GPS/CSOs, and we therefore investigated the dependence of expansion velocity on radio luminosity. We have searched the literature for detections of and upper limits to expansion velocities in CSOs. Since the first detections of lobe proper motions in \object{0710+439} (Owsianik \& Conway \cite{Owsianik98}) and \object{0108+388} (Owsianik, Conway \& Polatidis \cite{Owsianik98b}) expansion velocities have been measured for only a few other sources. In Table \ref{TCOR_4} we summarize these results, and also include upper limits to expansion velocities for which no significant expansion has been measured so far. Col. 1 gives the source IAU name, Col. 2 the redshift, Col. 3 the lobe-to-lobe, projected expansion velocity in units of $h^{-1} c$, converted to the adopted cosmology (WMAP5, see Sect. \ref{COR_int}), Col. 4 the number of observing epochs and the years of the first and last observations, and Col. 5 gives the reference for the expansion velocity. Note that the upper limits as presented in Gugliucci et al. (\cite{Gugliucci05}) had not been corrected for cosmological time dilation. The values we present here have been corrected and represent rest-frame quantities.

In Fig. \ref{FCOR_vL} we show the expansion velocities measured in GPS/CSOs as a function of radio luminosity. Both the radio luminosity and the lobe-to-lobe, projected expansion velocity have been corrected for cosmological effects, and represent rest-frame quantities. The upper limit indicated with the cross is our 3$\sigma$ upper limit to the expansion velocity of \object{CORALZ J083139+460800}, the other arrows indicate 1$\sigma$ upper limits from the literature (see Table \ref{TCOR_4}).

Although there is a large scatter, a positive correlation with radio luminosity can be seen. 
A Spearman rank correlation analysis shows that, ignoring all measured upper limits to the expansion velocity, the correlation is significant at a 98\% confidence level ($n=9$, $\rho = 0.717$, one-tailed). With the upper limits included in the analysis, the correlation reaches confidence levels $>98\%$, regardless of the precise way we treat these upper limits.
Furthermore it seems that the scatter can mostly be explained in terms of projection effects, since the majority of the sources lie between the solid and dashed lines, comprising a range of 90\% probability.

Note that we do expect a correlation between expansion velocity and radio luminosity from radio source evolution theory. In the appendix in Sect. \ref{COR_model} we present an analytic model of a young, expanding radio source with a constant jet power in a power-law density environment. This model predicts that $v_{\rm exp} \propto L_{\rm radio}^{1/3}$. The solid line represents such a relation, with arbitrary scaling. The dashed line represents this relation scaled down by a factor of $\sqrt{1-0.9^2} \simeq 0.44$, such that 90\% of a sample of randomly oriented sources that intrinsically would follow the solid line, should be located above the dashed line.

\begin{table}
\caption{Expansion velocities of Compact Symmetric Objects\label{TCOR_4} presented in the literature.}
\begin{center}
\begin{tabular}{llccc}
 \hline
 IAU Name & \multicolumn{1}{c}{$z$} & $v_{\rm exp}$ & No. of epochs & Ref. \\
          &                       & $h^{-1} c$ \\
 \hline
 0035+227       & 0.096  & 0.13$\pm$0.06 & 2 (1998-2001) & 1 \\
 0108+388       & 0.669  & 0.23$\pm$0.01 & 5 (1982-2000) & 1 \\
 0710+439       & 0.518  & 0.36$\pm$0.02 & 7 (1980-2000) & 1 \\
 1031+567       & 0.460  & 0.29$\pm$0.10 & 2 (1995-1999) & 2 \\
 OQ208          & 0.077  & 0.21$\pm$0.08 & 6 (1993-1997) & 3 \\
 1843+356       & 0.763  & 0.51$\pm$0.05 & 2 (1993-1997) & 1 \\
 1943+546       & 0.263  & 0.29$\pm$0.04 & 4 (1993-2000) & 1 \\
 2021+614       & 0.227  & 0.13$\pm$0.02 & 3 (1982-1998) & 4 \\
 2352+495       & 0.238  & 0.13$\pm$0.03 & 6 (1983-2000) & 1 \\

 J1111+1955     & 0.299  & $<$0.13       & 3 (1997-2002) & 5 \\
 J1414+4554     & 0.190  & $<$0.12       & 3 (1997-2002) & 5 \\
 1718-649       & 0.014  & $<$0.06       & 2 (1993-1999) & 6 \\
 J1734+0926     & 0.735  & $<$0.23       & 3 (1997-2002) & 5 \\
 1934-638       & 0.183  & $<$0.25       & 3 (1969-1988) & 7 \\
 \hline
\end{tabular}
\end{center}
References: (1) Polatidis \& Conway (\cite{Polatidis03}); (2) Taylor et al. (\cite{Taylor00}); (3) Stanghellini et al. (\cite{Stanghellini02}); (4) Tschager et al. (\cite{Tschager00}); (5) Gugliucci et al. (\cite{Gugliucci05}); (6) Tingay et al. (\cite{Tingay02}); (7) Tzioumis et al. (\cite{Tzioumis98}).
\end{table}

\section{Conclusions} \label{COR_con}

In this paper we present VLBI and MERLIN observations of the CORALZ sample, a nearby sample of 25 compact ($\theta<2''$) radio sources, of which 18 form a 95\% statistically complete sample in the redshift range $0.005<z<0.16$. We have measured the sizes of the sources, and performed a morphological classification. 

The CORALZ core sample follows the well established relation between radio spectral peak frequency and angular size, but with significantly smaller sizes at any particular peak frequency, compared to more powerful and more distant GPS/CSS sources. We show that this is exactly as expected from synchrotron self-absorption theory, in which the angular size is also proportional to the square root of the peak flux density. By least-squares fitting of the combined sample of CORALZ and bright literature samples, we recover the dependencies of angular size on peak frequency and peak flux density as expected for synchrotron self-absorbed radio sources evolving in a self-similar way. Current models that invoke FFA to explain the spectral turnovers in GPS and CSS sources (Bicknell et al. \cite{Bicknell97}) can not explain the relatively small angular sizes found for the CORALZ core sample. We therefore conclude that, although FFA may play a role in some sources, the turnovers must be caused by SSA for the majority of GPS and CSS radio sources.

In addition, based on two epochs of global VLBI, we have derived a strong upper limit to the expansion velocity of \object{CORALZ J083139+460800} ($v_{exp} <$ 0.07 $h^{-1} c$). We show that, in comparison with more powerful young radio sources, the expansion velocity is low, indicating that the expansion speed of GPS radio sources is correlated to their radio luminosity. We show that this is expected from analytic radio source evolution modelling, which predicts $v_{exp} \propto L_{radio}^{1/3}$ (appendix Sec. \ref{COR_model}). The scatter in the correlation is consistent with being due to projection effects.

\begin{acknowledgements}
The European VLBI Network is a joint facility of European, Chinese, South African and other radio astronomy institutes funded by their national research councils
(proposal codes: ES042, GS016, GS021).
The National Radio Astronomy Observatory is a facility of the National Science Foundation operated under cooperative agreement by Associated Universities, Inc
(proposal codes: GS016, GS021).
MERLIN is a National Facility operated by the University of Manchester at Jodrell Bank Observatory on behalf of STFC
(proposal codes: MN/00B/07, ES042).
This research has made use of the NASA/IPAC Extragalactic Database (NED) which is operated by the Jet Propulsion Laboratory, California Institute of Technology, under contract with the National Aeronautics and Space Administration.
\end{acknowledgements}

\clearpage

\appendix

\section{Radio source evolution modelling} \label{COR_model}
In this section we present analytic models predicting how the expansion velocity of young radio sources depends on radio luminosity $L_{\rm radio}$, source age $t$, and environment. Our consideration is based on the radio source evolution models of Kaiser \& Best (\cite{Kaiser07}), hereafter KB. If the sources are physically small (less than a few kpc in size) and their radio luminosity is measured at a frequency where self-absorption is not important, they are in the `synchrotron loss dominated' regime as discussed in KB (Sect. 2.3). In this regime we expect the radio luminosity $L_{\rm radio}$ to be proportional to $Q$, the jet power, which is assumed to be constant with time, and independent of the density of the source environment.
If we assume that the density $\rho$ in the source environment follows a power-law distribution with respect to $r$, the distance to the centre, then
\begin{equation}
 \rho = \rho_0 \left( \frac{r}{a_0} \right)^{-\beta},
\end{equation}
where $\rho_0$ is the density at an arbitrary reference point $a_0$, and $\beta$ is the slope of the density profile. Note that $\rho_0$ and $a_0$ are not independent, and the model can only depend on their combination $\rho_0 a_0^\beta$.
Using equation A2 from the appendix of KB, the linear size of the source, $D$, is given by
\begin{equation}
 D \propto \left( \frac{Q}{\rho_0 a_0^{\beta}} \right)^{1 / \left( 5 - \beta \right)} t^{3/ \left( 5 - \beta \right)}.
\end{equation}
This means that the expansion speed of the lobe is given by
\begin{equation}
v_{\rm exp} = \dot{D} \propto Q^{1/\left(5-\beta\right)} t^{\left( \beta - 2 \right) / \left( 5 - \beta \right)} \left( \rho_0 a_0^{\beta} \right)^{-1/\left( 5 - \beta \right)}.
\end{equation}
Assuming $Q$ is proportional to $L_{\rm radio}$, we obtain
\begin{equation} \label{COR_eqVexp}
v_{\rm exp} \propto L_{\rm radio}^{1/\left( 5 - \beta \right)} t^{\left( \beta - 2 \right) / \left( 5 - \beta \right)} \left( \rho_0 a_0^{\beta} \right)^{-1 / \left( 5 - \beta \right)}.
\end{equation}
If we take the often adopted value $\beta =2$ (e.g. Bicknell et al. \cite{Bicknell97}), the time dependence drops out and equation (\ref{COR_eqVexp}) reduces to
\begin{equation}
v_{\rm exp} \propto L_{\rm radio}^{1/3} \left( \rho_0 a_0^{2} \right)^{-1/3}.
\end{equation}
We have used this equation for comparison with the observed expansion speeds in Sect. \ref{COR_VexpL}. Note that in the central part of galaxies one might expect the density profile to become less steep, even possibly resulting in a flat density profile with $\beta = 0$. In this extreme case you would expect $v_{\rm exp} \propto L_{\rm radio}^{1/5} t^{-2/5} \rho_0^{-1/5}$, so the slope of the correlation between expansion velocity and radio power is not extremely sensitive to the value of $\beta$, but the additional time dependence may increase the scatter around it, in addition to the scatter caused by projection effects and by variations in the density profile of the surrounding medium from source to source.

\begin{table*}[!h]
\section{Observations and radio maps}
\centering
\caption{Observations of the sources in the CORALZ sample. The first 18 objects form the CORALZ core sample at $0.005<z<0.160$, the next seven objects are other nearby radio sources in the CORALZ sample, and the final three objects are other compact radio sources located towards probably random foreground galaxies. \label{TCOR_1}}
\begin{tabular}{ccccccrcr@{ $\times$ }lr}
\hline
IAU Name & Fig. & Epoch & Freq. & Bandw. & Array & $S_{\rm{peak}}$                       &
 rms                       & \multicolumn{3}{c}{Beam} \\
         &      &       & (GHz) & (MHz)  &       & \multicolumn{2}{c}{(mJy beam$^{-1}$)} &
 \multicolumn{2}{c}{(mas)} & $\degr$                  \\
\hline
\multicolumn{8}{l}{The CORALZ core sample} \\
\object{J073328+560541} & \ref{FCOR_1} & 2000/02/14 & 1.655 & 32 & EVN      & 126 & 0.39 &  25 &  15 & 31 \\
                        & \ref{FCOR_1} & 2004/06/05 & 1.665 & 16 & EVN+VLBA &  55 & 0.22 & 7.8 & 2.1 & -12 \\
                        & \ref{FCOR_1} & 2004/05/24 & 4.993 & 16 & EVN+VLBA &  47 & 0.19 & 9.0 & 7.9 & -59 \\
\object{J073934+495438} & \ref{FCOR_2} & 2000/03/02 & 4.987 & 32 & EVN+VLBA &  22 & 0.06 & 2.8 & 0.8 & -25 \\
                        & \ref{FCOR_2} & 2004/05/24 & 4.993 & 16 & EVN+VLBA &  25 & 0.06 & 4.1 & 1.3 & -32 \\
\object{J083139+460800} & \ref{FCOR_2} & 2000/03/02 & 4.987 & 32 & EVN+VLBA &  32 & 0.13 & 2.3 & 0.7 & -6 \\
                        & \ref{FCOR_2} & 2004/05/24 & 4.993 & 16 & EVN+VLBA &  42 & 0.08 & 3.1 & 0.8 & -13 \\
\object{J083637+440109} & \ref{FCOR_3} & 2000/11/13 & 1.658 & 15 & MERLIN   &  22 & 0.09 & 125 & 121 & -75 \\
\object{J090615+463618} & \ref{FCOR_1} & 2001/05/27 & 1.659 & 16 & EVN      & 134 & 0.13 &  22 &  13 & 59 \\
                        & \ref{FCOR_1} & 2004/06/05 & 1.665 & 16 & EVN+VLBA &  88 & 0.16 & 9.9 & 2.1 & -4 \\
                        & \ref{FCOR_1} & 2004/05/24 & 4.993 & 16 & EVN+VLBA &  84 & 0.18 & 2.8 & 0.7 & -14 \\
\object{J102618+454229} & \ref{FCOR_1} & 2001/05/28 & 1.659 & 16 & EVN      &  76 & 0.14 &  22 &  15 & 69 \\
                        & \ref{FCOR_1} & 2004/06/05 & 1.665 & 16 & EVN+VLBA &  30 & 0.10 &  10 & 2.1 & -3 \\
                        & \ref{FCOR_1} & 2004/05/24 & 4.993 & 16 & EVN+VLBA &  30 & 0.06 & 2.7 & 0.8 & -3 \\
\object{J103719+433515} & \ref{FCOR_1} & 2001/05/27 & 1.659 & 16 & EVN      &  61 & 0.23 &  21 &  13 & 77 \\
                        & \ref{FCOR_1} & 2004/06/05 & 1.665 & 16 & EVN+VLBA &  21 & 0.10 &  12 & 2.1 & -7 \\
                        & \ref{FCOR_1} & 2004/05/24 & 4.993 & 16 & EVN+VLBA &  19 & 0.23 &  11 & 6.7 & 34 \\
\object{J115000+552821} & \ref{FCOR_3} & 2001/05/28 & 1.659 & 16 & EVN      & 103 & 0.14 &  20 &  14 & 63 \\
\object{J120902+411559} & \ref{FCOR_1} & 2001/05/27 & 1.659 & 16 & EVN      & 117 & 0.17 &  23 &  14 & 48 \\
                        & \ref{FCOR_1} & 2004/06/05 & 1.665 & 16 & EVN+VLBA &  54 & 0.05 & 9.5 & 2.1 & -11 \\
                        & \ref{FCOR_1} & 2004/05/24 & 4.993 & 16 & EVN+VLBA &  73 & 0.54 & 9.1 & 6.1 & 43 \\
\object{J131739+411545} & \ref{FCOR_2} & 2000/03/02 & 4.987 & 32 & EVN+VLBA &  45 & 0.22 & 1.4 & 0.9 & -25 \\
                        & \ref{FCOR_2} & 2004/05/24 & 4.993 & 16 & EVN+VLBA &  53 & 0.11 & 2.7 & 0.7 & -3 \\
\object{J140051+521606} & \ref{FCOR_3} & 2000/12/26 & 1.408 & 15 & MERLIN   & 101 & 0.20 & 163 & 150 & -82 \\
\object{J140942+360416} & \ref{FCOR_1} & 2001/05/28 & 1.659 & 16 & EVN      &  81 & 0.13 &  28 &  11 & 37 \\
                        & \ref{FCOR_1} & 2004/06/05 & 1.665 & 16 & EVN+VLBA &  30 & 0.06 &  17 & 2.1 & -5 \\
\object{J143521+505122} & \ref{FCOR_3} & 2000/12/26 & 1.408 & 15 & MERLIN   &  73 & 0.26 & 166 & 146 & -41 \\
\object{J150805+342323} & \ref{FCOR_3} & 2002/02/19 & 1.650 & 16 & EVN      &   4 & 0.11 &  24 &  23 & 16 \\
\object{J160246+524358} & \ref{FCOR_1} & 2001/05/27 & 1.659 & 16 & EVN      &  62 & 0.14 &  21 &  12 & 59 \\
                        & \ref{FCOR_1} & 2004/06/05 & 1.665 & 16 & EVN+VLBA &  23 & 0.16 & 8.4 & 2.1 & -19 \\
                        & \ref{FCOR_1} & 2004/05/24 & 4.993 & 16 & EVN+VLBA &  30 & 0.13 & 7.6 & 6.3 & 58 \\
\object{J161148+404020} & \ref{FCOR_3} & 2000/11/25 & 1.658 & 15 & MERLIN   &  86 & 0.12 & 167 & 136 & -1 \\
\object{J170330+454047} \\
\object{J171854+544148} & \ref{FCOR_1} & 2000/02/14 & 1.655 & 32 & EVN      & 119 & 0.33 &  20 &  13 & -84 \\
                        & \ref{FCOR_1} & 2004/06/05 & 1.665 & 16 & EVN+VLBA &  90 & 0.19 &  10 & 2.1 & -6 \\
                        & \ref{FCOR_1} & 2004/05/24 & 4.993 & 16 & EVN+VLBA &  49 & 0.11 & 8.9 & 6.1 & 16 \\
\multicolumn{8}{l}{Other nearby sources in the CORALZ sample} \\
\object{J093609+331308} & \ref{FCOR_4} & 2000/03/02 & 4.987 & 32 & EVN+VLBA &  28 & 0.15 & 1.2 & 0.8 & 0 \\
\object{J101636+563926} & \ref{FCOR_4} & 2000/11/09 & 1.658 & 15 & MERLIN   &  52 & 0.08 & 174 & 136 & -31 \\
\object{J105731+405646} & \ref{FCOR_4} & 2000/03/02 & 4.987 & 32 & EVN+VLBA &  14 & 0.08 & 1.8 & 0.3 & 13 \\
\object{J115727+431806} & \ref{FCOR_4} & 2001/01/02 & 1.408 & 15 & MERLIN   &  41 & 0.09 & 182 & 173 & -47 \\
\object{J132513+395552} & \ref{FCOR_4} & 2000/03/02 & 4.987 & 32 & EVN+VLBA &  15 & 0.08 & 0.7 & 0.3 & -20 \\
\object{J134035+444817} & \ref{FCOR_4} & 2000/03/02 & 4.987 & 32 & EVN+VLBA &  76 & 0.13 & 1.4 & 0.9 & -7 \\
\object{J155927+533054} & \ref{FCOR_4} & 2000/12/18 & 1.408 & 15 & MERLIN   &  16 & 0.11 & 308 & 202 & 14 \\
\hline
\multicolumn{8}{l}{Additional sources} \\
\object{J071509+452555} & \ref{FCOR_5} & 2000/03/02 & 4.987 & 32 & EVN+VLBA &  30 & 0.13 & 0.8 & 0.8 &  0 \\
\object{J080454+433537} & \ref{FCOR_5} & 2000/02/14 & 1.655 & 32 & EVN      & 182 & 0.39 &  27 &  14 & 39 \\
\object{J134158+541524} & \ref{FCOR_5} & 2002/02/19 & 1.650 & 16 & EVN      &  10 & 0.13 &  25 &  22 & 70 \\
\hline
\end{tabular}
\end{table*}

\begin{figure*}[ht]
  \centering
  \includegraphics[width=\textwidth]{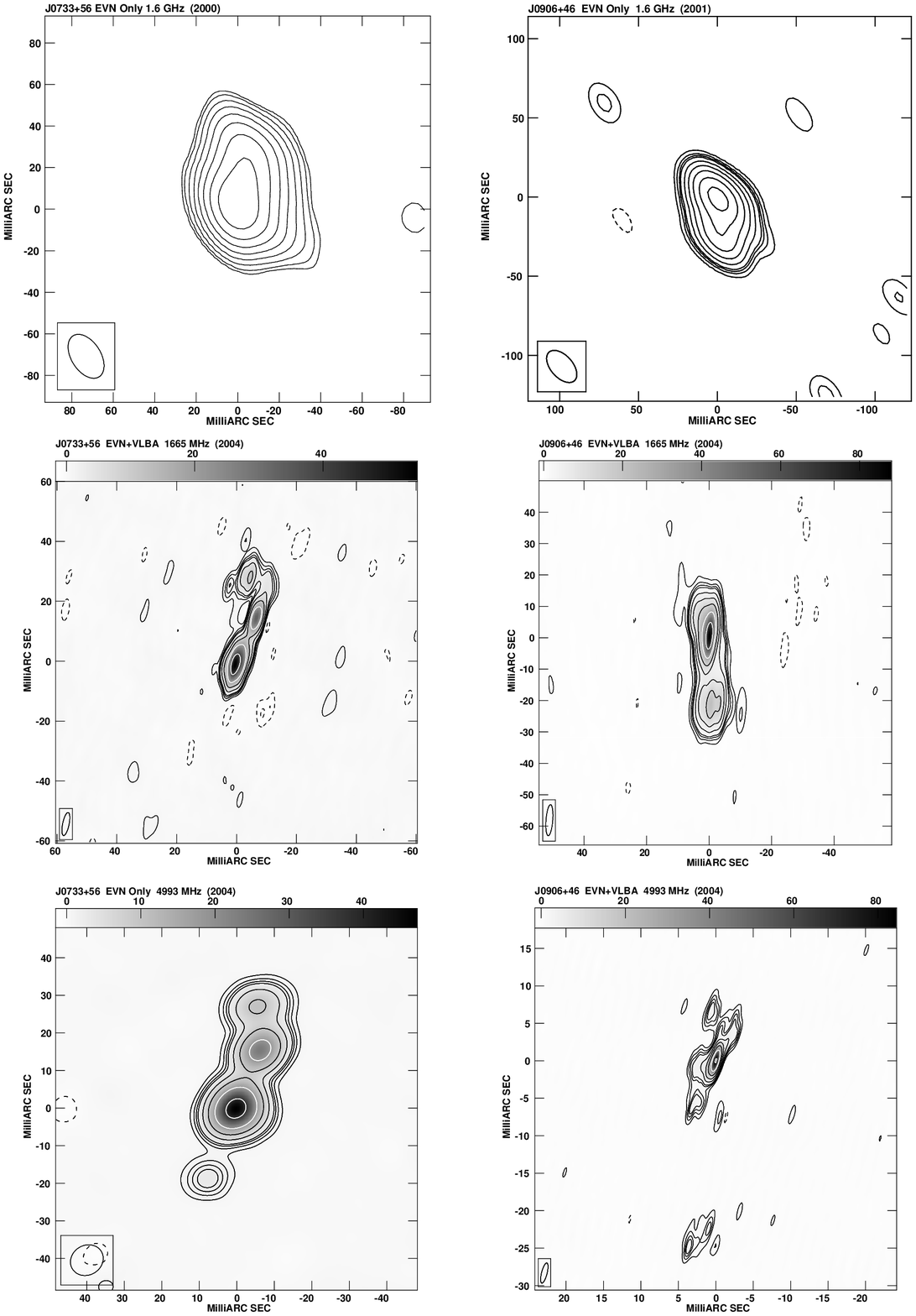}
  \caption{Contour maps of sources from the CORALZ core sample observed with the EVN (1.6 GHz, 2000; first row), and global VLBI (1.6 GHz and 5.0 GHz, 2004; second and third row, respectively).}
  \label{FCOR_1}
\end{figure*}
\clearpage
\addtocounter{figure}{-1}
\begin{figure*}[ht]
  \centering
  \includegraphics[width=\textwidth]{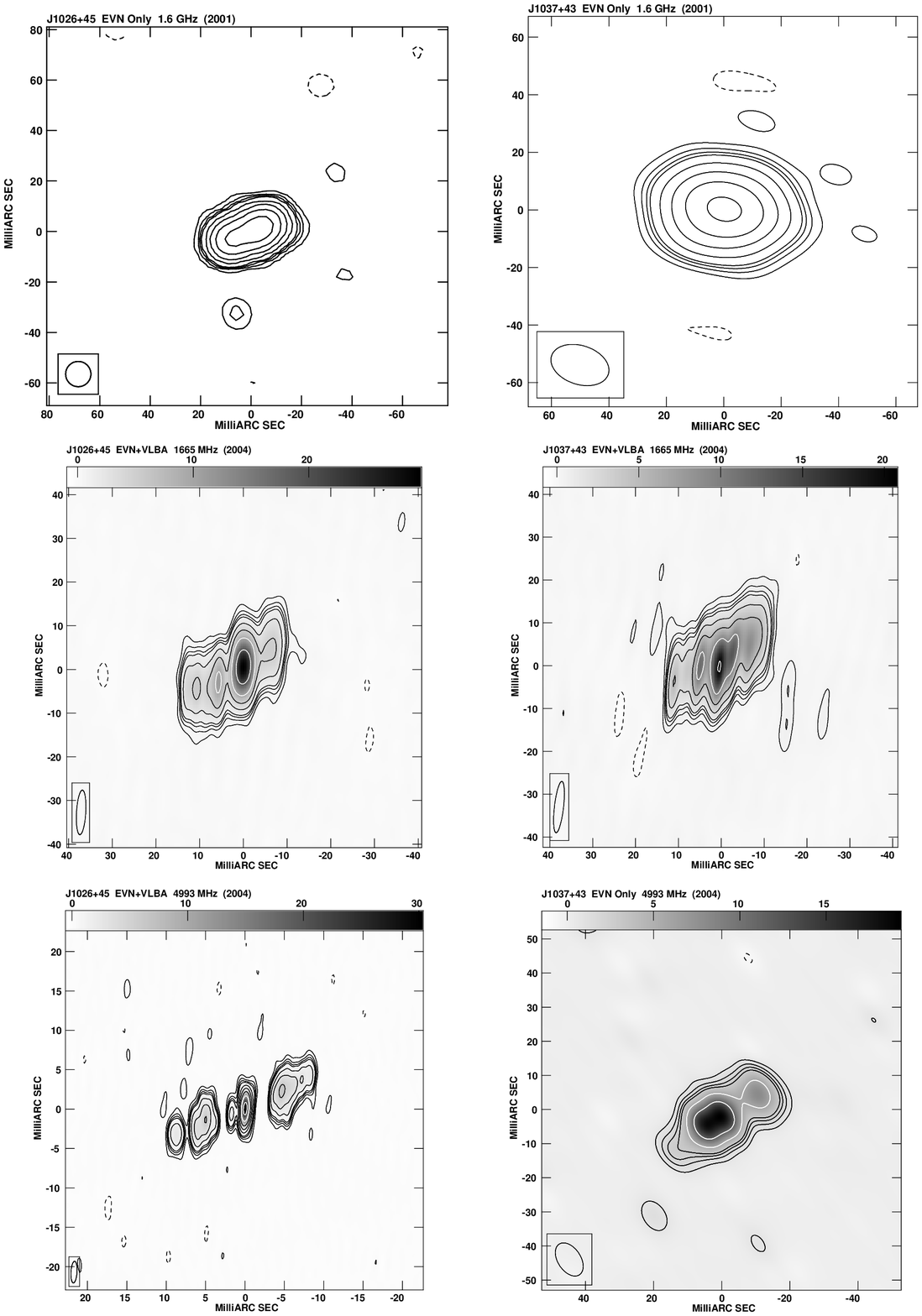}
  \caption{continued.}
\end{figure*}
\clearpage
\addtocounter{figure}{-1}
\begin{figure*}[ht]
  \centering
  \includegraphics[width=\textwidth]{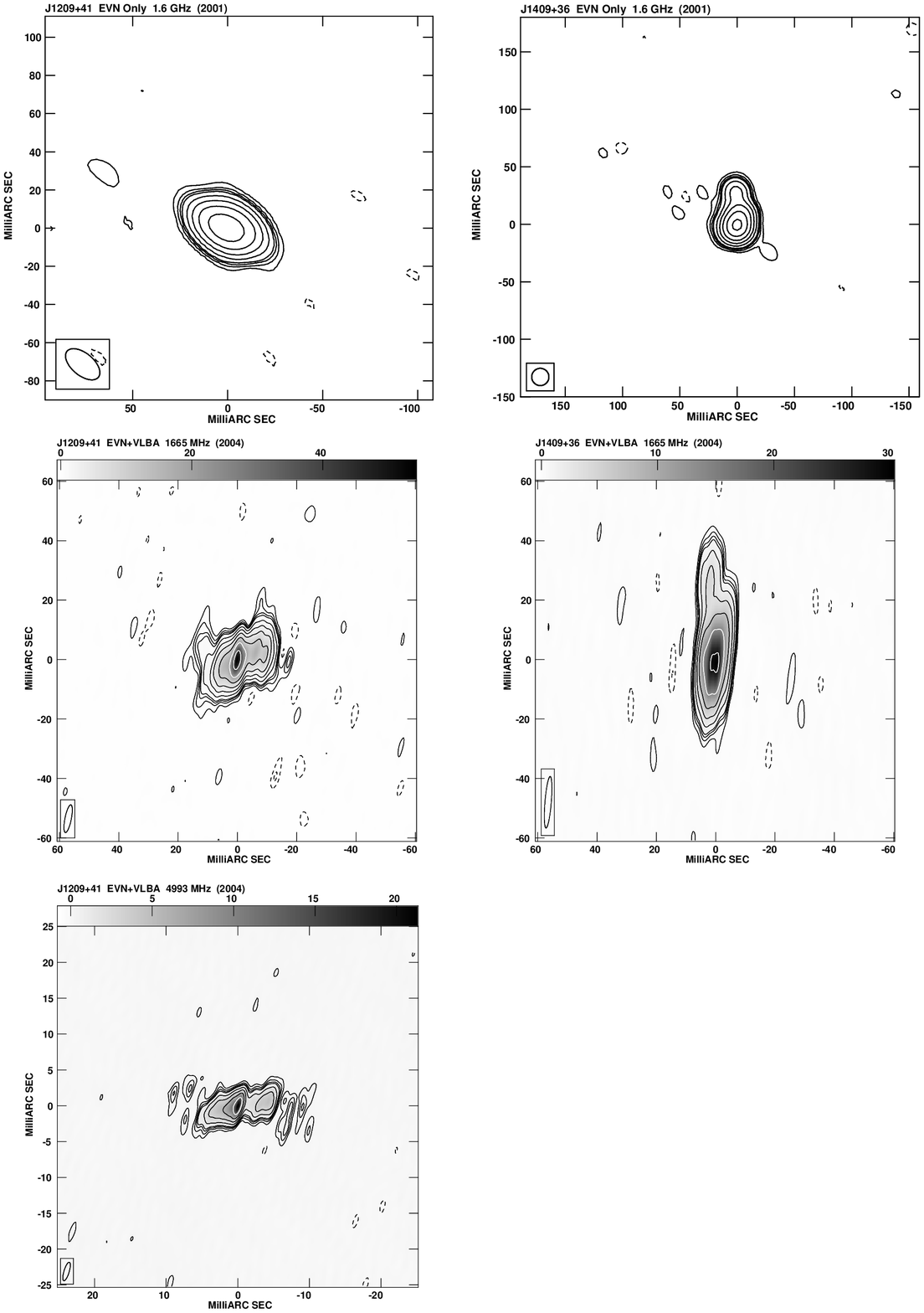}
  \caption{continued.}
\end{figure*}
\clearpage
\addtocounter{figure}{-1}
\begin{figure*}[ht]
  \centering
  \includegraphics[width=\textwidth]{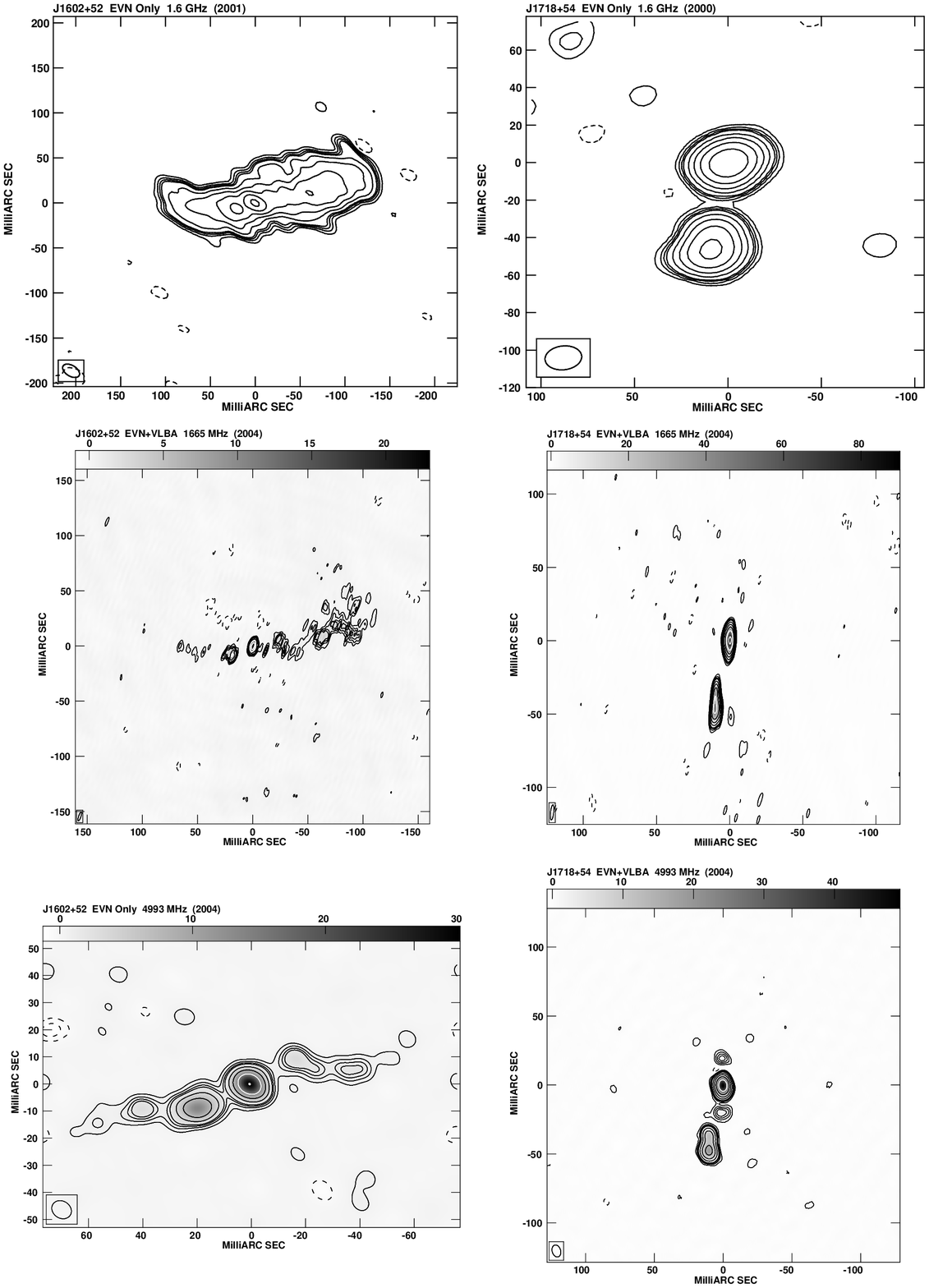}
  \caption{continued.}
\end{figure*}
\clearpage
\begin{figure*}[ht]
  \centering
  \includegraphics[width=0.9\textwidth]{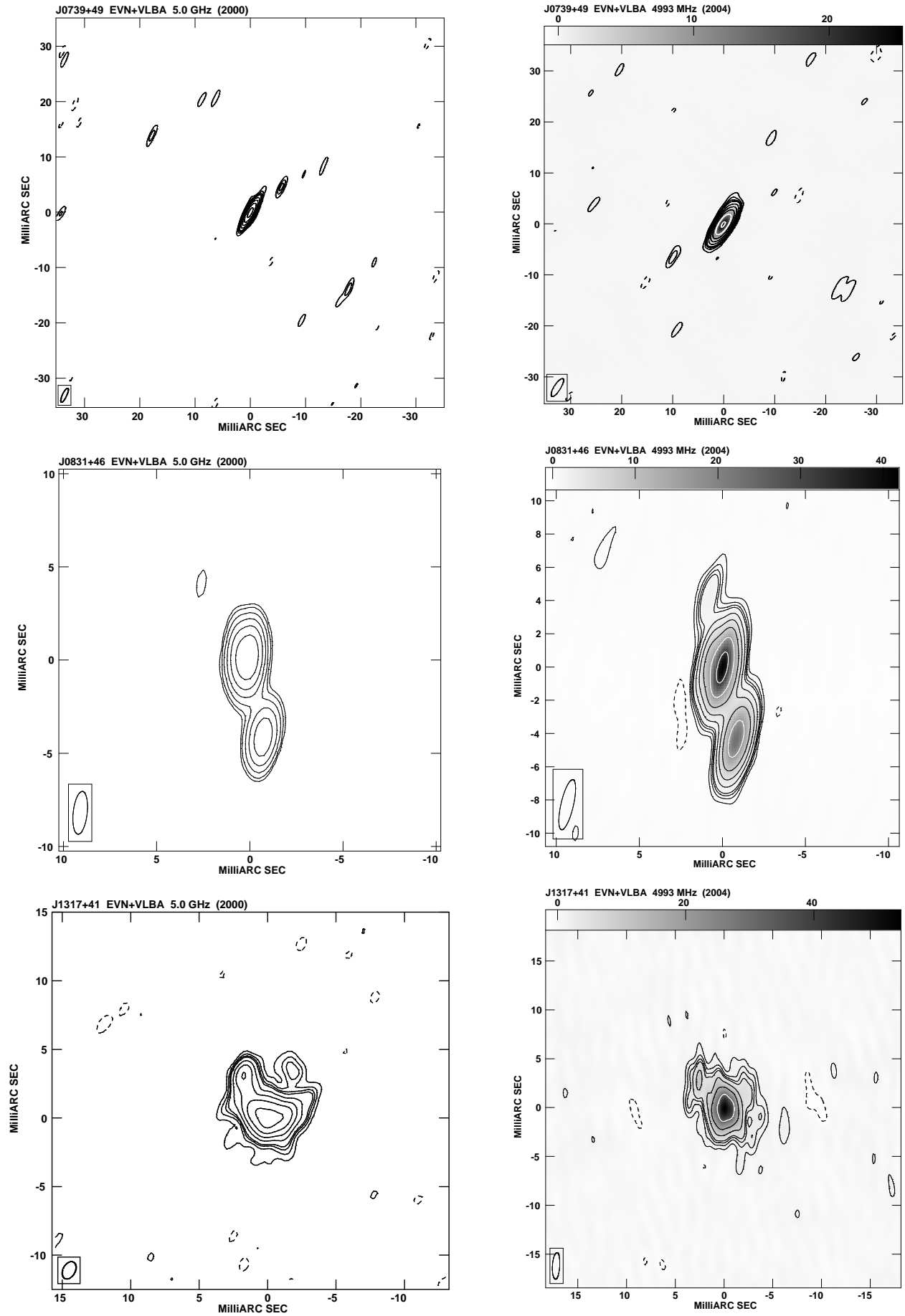}
 \caption{Contour maps of the three most compact sources from the CORALZ core sample taken with global VLBI at 5.0 GHz in 2000 (first column), and 2004 (second column).}
  \label{FCOR_2}
\end{figure*}
\clearpage
\begin{figure*}[ht]
  \centering
  \includegraphics[width=\textwidth]{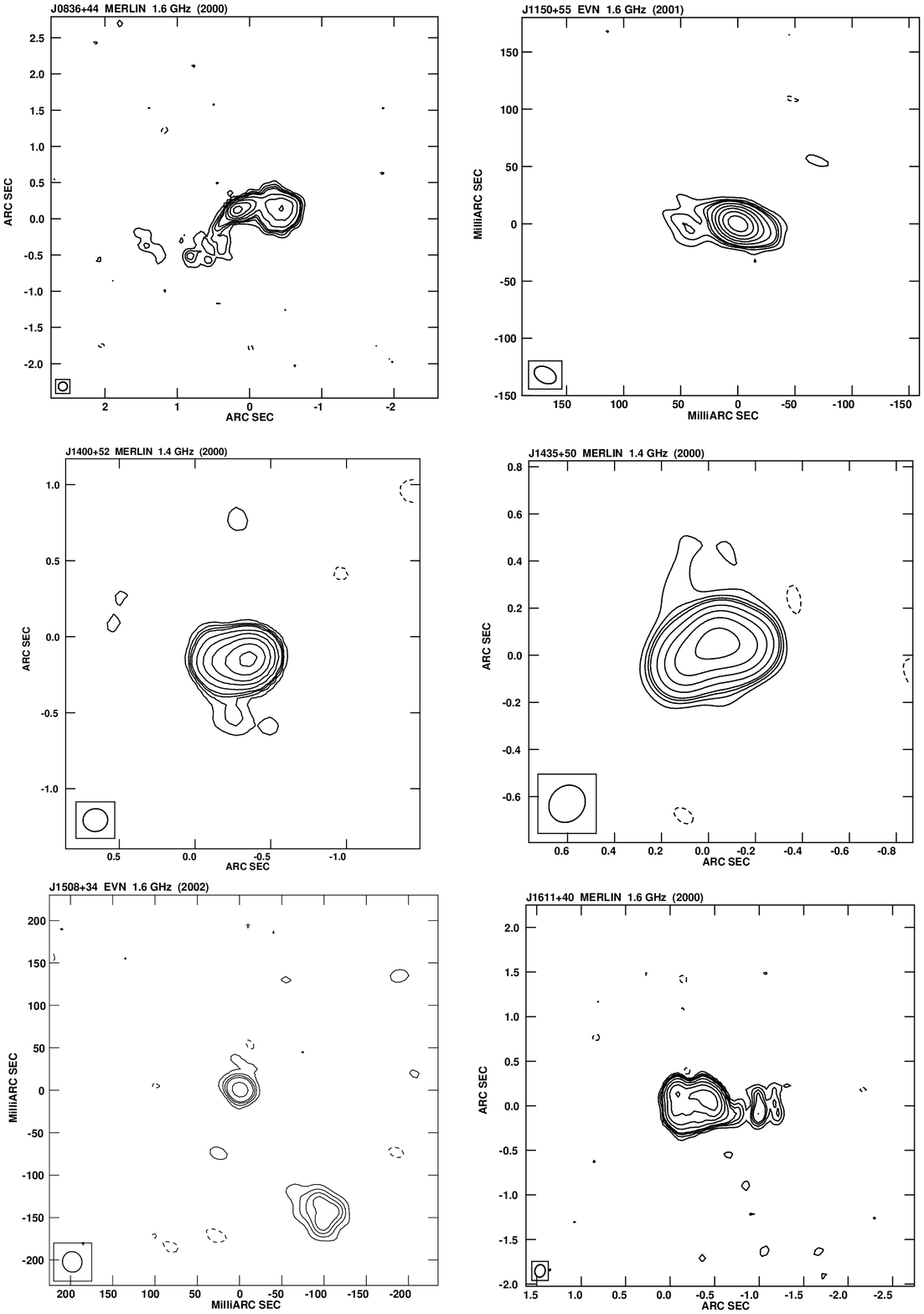}
  \caption{Contour maps of the more extended sources from the CORALZ core sample taken with MERLIN at 1.4 GHz or 1.6 GHz in 2000, or with the EVN at 1.6 GHz.}
  \label{FCOR_3}
\end{figure*}
\clearpage
\begin{figure*}[ht]
  \centering
  \includegraphics[width=\textwidth]{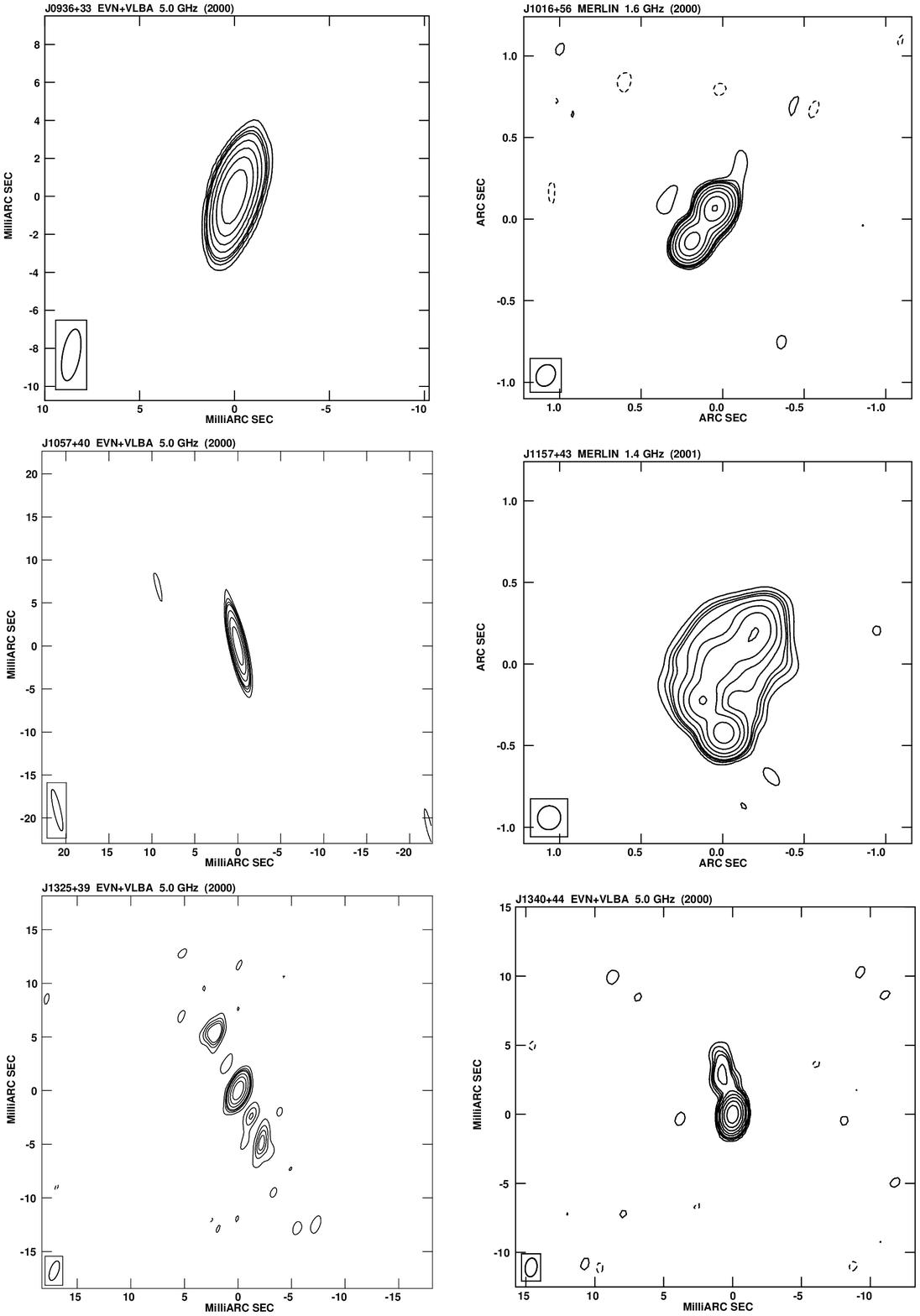}
  \caption{Contour maps of other sources from the CORALZ sample taken in 2000 with global VLBI (5.0 GHz) or with MERLIN (1.4 GHz or 1.6 GHz). These sources are either slightly too faint, or too distant to be part of the CORALZ core sample.}
  \label{FCOR_4}
\end{figure*}
\clearpage
\addtocounter{figure}{-1}
\begin{figure*}[ht]
  \includegraphics[width=8.8cm]{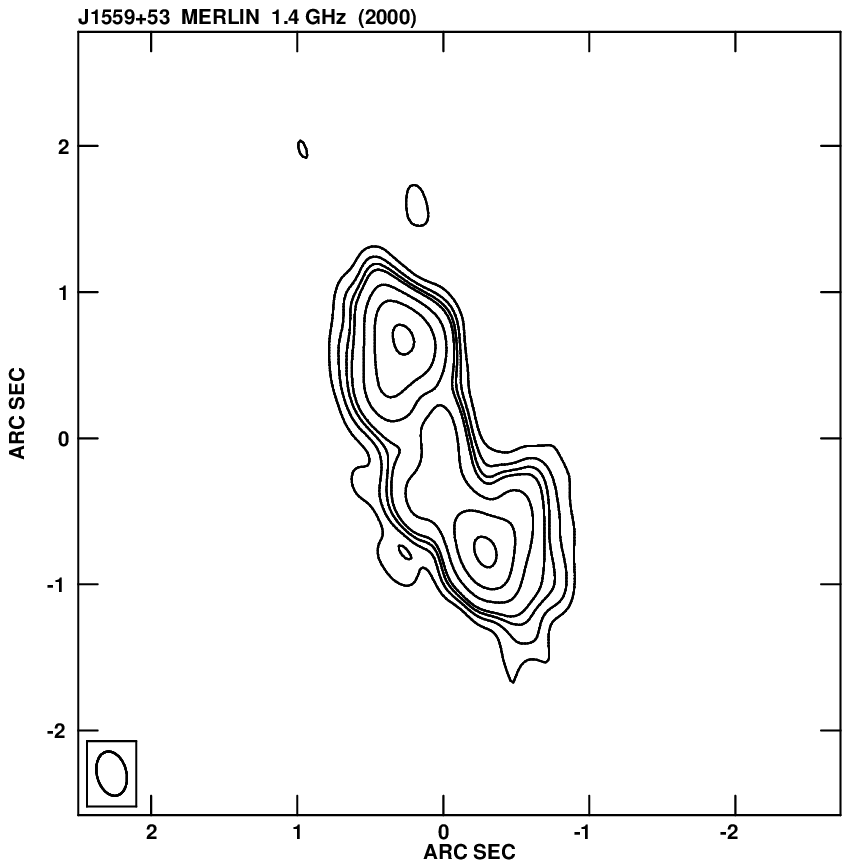}
  \caption{continued.}
\end{figure*}
\begin{figure*}[ht]
  \centering
  \includegraphics[width=\textwidth]{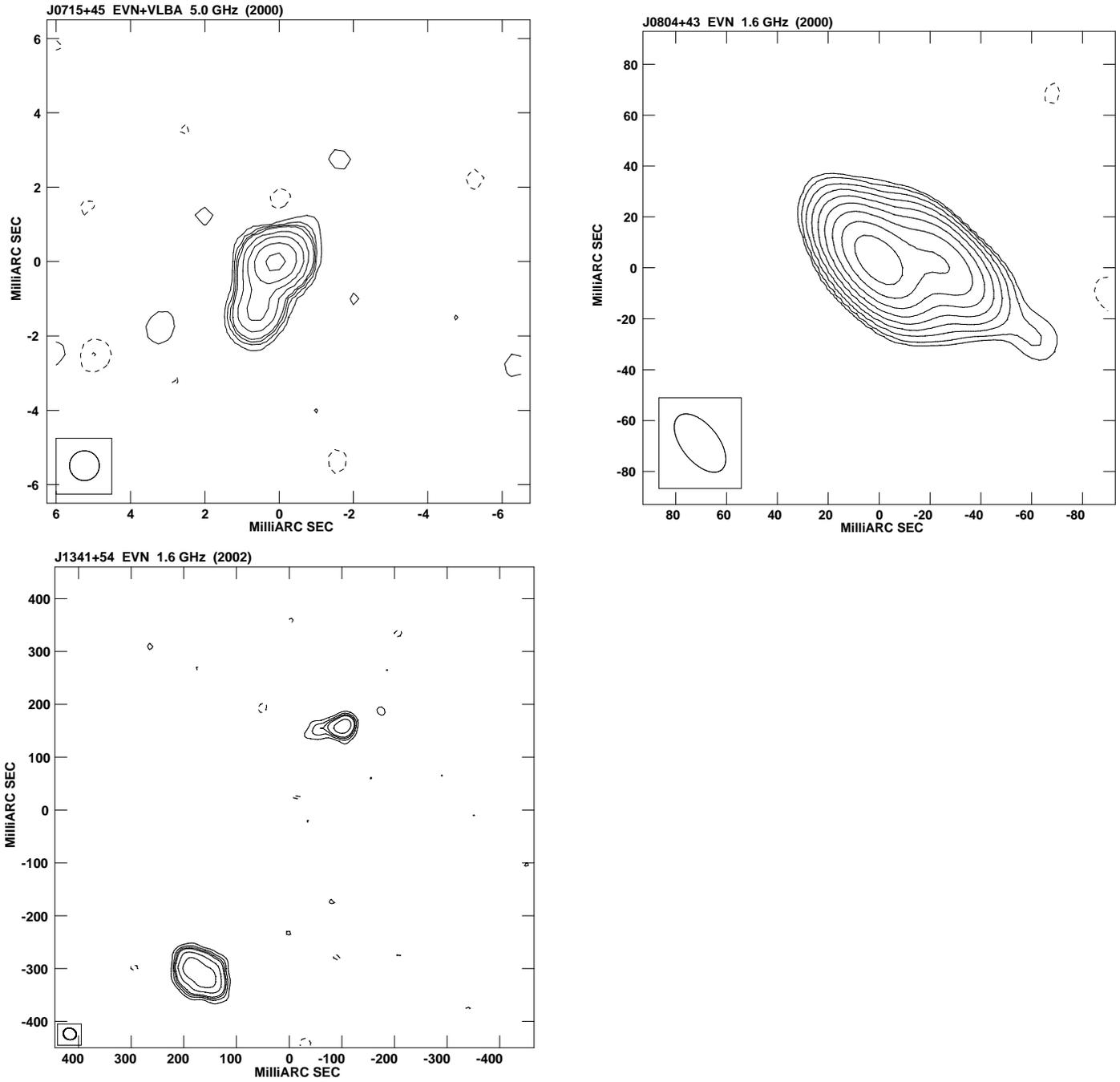}
  \caption{Contour maps of the additional sources, observed with global VLBI (5.0 GHz), or with the EVN (1.6 GHz). These sources have been excluded from the CORALZ sample because the offset between the radio and optical position is large, indicating that these radio sources are located towards probably random foreground galaxies.}
  \label{FCOR_5}
\end{figure*}

\begin{longtable}{c c c c c c r @{$\pm$} l r @{$\times$} l c c}
\caption{The fitted parameters of the observed components.\label{TCOR_3}}\\
\hline
IAU name & Fig & Epoch & $\nu_{\rm{obs}}$ & $\Delta x$ & $\Delta y$ & \multicolumn{2}{c}{Flux density}  & \multicolumn{2}{c}{Size}  & PA      \\
         &     &       & (GHz)       & (mas)      & (mas)      & \multicolumn{2}{c}{(mJy)} & \multicolumn{2}{c}{(mas)} & $\degr$ \\
\hline
\endfirsthead
\caption{The fitted parameters of the observed components, continued.}\\
\hline
IAU name & Fig & Epoch & $\nu_{\rm{obs}}$ & $\Delta x$ & $\Delta y$ & \multicolumn{2}{c}{Flux density}  & \multicolumn{2}{c}{Size}  & PA      \\
         &     &       & (GHz)       & (mas)      & (mas)      & \multicolumn{2}{c}{(mJy)} & \multicolumn{2}{c}{(mas)} & $\degr$ \\
\hline
\endhead
\hline
\endfoot
\object{J073328+560541} & \ref{FCOR_1} & 2000/02/14 & 1.655 &    0.0 &    0.0 &  248.0 &    0.9 &   29.4 &    8.4 &  164 \\
 & \ref{FCOR_1} & 2004/06/05 & 1.665 &    0.0 &    0.0 &  150.0 &    0.7 &    7.1 &    3.8 &  156 \\
& & & &   -6.7 &   15.3 &   60.1 &    0.5 &    5.5 &    2.2 &  153 \\
& & & &   -4.3 &   28.9 &   28.6 &    0.7 &    5.9 &    3.7 &  107 \\
 & \ref{FCOR_1} & 2004/05/24 & 4.993 &    0.0 &    0.0 &   58.9 &    0.1 &    5.1 &    3.4 &  147 \\
& & & &   -6.4 &   15.5 &   28.1 &    0.1 &    5.3 &    1.8 &  147 \\
& & & &   -5.8 &   26.8 &   14.8 &    0.1 &    6.3 &    4.8 &   80 \\
& & & &    7.5 &  -18.6 &    3.1 &    0.1 &    3.1 &    0.0 &   83 \\
\object{J073934+495438} & \ref{FCOR_2} & 2000/03/02 & 4.987 &    0.0 &    0.0 &   30.0 &    0.1 &    1.8 &    0.4 &  142 \\
 & \ref{FCOR_2} & 2004/05/24 & 4.993 &    0.0 &    0.0 &   36.7 &    0.1 &    2.0 &    1.0 &  140 \\
\object{J083139+460800} & \ref{FCOR_2} & 2000/03/02 & 4.987 &    0.0 &    0.0 &   54.1 &    0.2 &    1.0 &    0.8 &   52 \\
& & & &   -0.9 &   -4.3 &   19.9 &    0.2 &    0.8 &    0.1 &  106 \\
 & \ref{FCOR_2} & 2004/05/24 & 4.993 &    0.0 &    0.0 &   65.0 &    0.2 &    1.1 &    0.7 &   45 \\
& & & &   -0.9 &   -4.3 &   35.5 &    0.2 &    0.9 &    0.6 &   78 \\
& & & &    0.8 &    4.2 &    2.3 &    0.2 &    0.6 &    0.0 &  163 \\
\object{J083637+440109} & \ref{FCOR_3} & 2000/11/13 & 1.658 &    0.0 &    0.0 &   60.4 &    0.8 &  385.9 &  295.2 &   76 \\
& & & &  546.3 &   -5.4 &   46.7 &    0.3 &  212.6 &   64.4 &  114 \\
\object{J090615+463618} & \ref{FCOR_1} & 2001/05/27 & 1.659 &    0.0 &    0.0 &  160.7 &    0.5 &    9.5 &    3.3 &    8 \\
& & & &   -1.3 &  -22.5 &   65.1 &    0.5 &   11.1 &    4.2 &    3 \\
 & \ref{FCOR_1} & 2004/06/05 & 1.665 &    0.0 &    0.0 &  160.7 &    0.5 &    9.5 &    3.3 &    8 \\
& & & &   -1.3 &  -22.5 &   65.1 &    0.5 &   11.1 &    4.2 &    3 \\
 & \ref{FCOR_1} & 2004/05/24 & 4.993 &    0.0 &    0.0 &  105.0 &    0.3 &    1.1 &    0.3 &  139 \\
& & & &    0.4 &    6.4 &    8.5 &    0.5 &    2.5 &    1.4 &  178 \\
& & & &    2.6 &   -5.5 &   10.3 &    0.6 &    2.7 &    1.6 &  150 \\
& & & &   -0.4 &   -7.4 &    2.2 &    0.4 &    2.8 &    0.2 &  171 \\
& & & &    3.4 &  -24.6 &    6.4 &    0.5 &    1.7 &    1.3 &  165 \\
\object{J102618+454229} & \ref{FCOR_1} & 2001/05/28 & 1.659 &    0.0 &    0.0 &  111.0 &    0.3 &   17.9 &    2.9 &  112 \\
 & \ref{FCOR_1} & 2004/06/05 & 1.665 &    0.0 &    0.0 &   59.0 &    0.2 &    3.5 &    1.8 &   84 \\
& & & &    5.7 &   -3.3 &    9.6 &    0.2 &    1.5 &    0.0 &  152 \\
& & & &   -6.4 &    4.1 &   13.8 &    0.3 &    5.8 &    3.9 &  173 \\
& & & &   10.4 &   -5.0 &   22.8 &    0.4 &    8.7 &    3.4 &   82 \\
 & \ref{FCOR_1} & 2004/05/24 & 4.993 &    0.0 &    0.0 &   33.7 &    0.1 &    0.8 &    0.3 &  173 \\
& & & &    5.3 &   -1.7 &   21.7 &    0.3 &    3.4 &    1.8 &  159 \\
& & & &   -4.8 &    2.2 &   26.0 &    0.3 &    3.0 &    1.8 &  146 \\
& & & &    8.8 &   -3.3 &    5.5 &    0.2 &    1.4 &    1.1 &   40 \\
& & & &    1.7 &   -0.6 &    2.2 &    0.1 &    0.7 &    0.5 &  164 \\
& & & &   -7.1 &    3.6 &    8.9 &    0.2 &    2.6 &    1.0 &  133 \\
\object{J103719+433515} & \ref{FCOR_1} & 2001/05/27 & 1.659 &    0.0 &    0.0 &   72.0 &    0.3 &   13.4 &    8.8 &  107 \\
 & \ref{FCOR_1} & 2004/06/05 & 1.665 &    0.0 &    0.0 &   24.7 &    0.2 &    3.0 &    1.6 &  175 \\
& & & &   -3.0 &    2.9 &   11.7 &    0.2 &    4.5 &    0.5 &  165 \\
& & & &    4.1 &    0.7 &   24.3 &    0.3 &    6.2 &    2.1 &  129 \\
& & & &   -7.2 &    6.0 &   25.1 &    0.3 &    6.2 &    4.8 &  140 \\
& & & &    9.3 &   -2.4 &   10.4 &    0.3 &    5.6 &    3.1 &  131 \\
 & \ref{FCOR_1} & 2004/05/24 & 4.993 &    0.0 &    0.0 &   21.0 &    0.1 &    4.2 &    1.8 &  131 \\
& & & &    5.5 &   -3.2 &   13.6 &    0.1 &    1.2 &    0.0 &   75 \\
& & & &  -10.3 &    6.3 &   12.9 &    0.1 &    7.1 &    5.2 &   91 \\
& & & &   11.4 &   -8.8 &    5.1 &    0.1 &    8.6 &    0.0 &  118 \\
\object{J115000+552821} & \ref{FCOR_3} & 2001/05/28 & 1.659 &    0.0 &    0.0 &  111.9 &    0.3 &    7.0 &    4.1 &   77 \\
& & & &   40.6 &    0.0 &    5.8 &    0.7 &   33.5 &   28.9 &   42 \\
\object{J120902+411559} & \ref{FCOR_1} & 2001/05/27 & 1.659 &    0.0 &    0.0 &  128.5 &    0.2 &    8.1 &    0.0 &   14 \\
 & \ref{FCOR_1} & 2004/06/05 & 1.665 &    0.0 &    0.0 &   56.3 &    0.1 &    2.6 &    0.0 &  131 \\
& & & &   -2.7 &    0.5 &   58.9 &    0.2 &   12.2 &    2.2 &  131 \\
& & & &   -8.6 &    1.0 &   25.1 &    0.2 &    4.8 &    0.0 &  171 \\
& & & &    4.5 &   -2.0 &   33.3 &    0.2 &    5.3 &    3.5 &  117 \\
 & \ref{FCOR_1} & 2004/05/24 & 4.993 &    0.0 &    0.0 &   19.7 &    0.1 &    0.6 &    0.0 &    8 \\
& & & &   -3.0 &    0.2 &    7.6 &    0.2 &    1.5 &    0.8 &  109 \\
& & & &   -4.5 &    0.8 &    9.3 &    0.2 &    1.7 &    1.2 &    1 \\
& & & &    1.4 &   -0.2 &   39.0 &    0.3 &    4.2 &    1.0 &  105 \\
\object{J131739+411545} & \ref{FCOR_2} & 2000/03/02 & 4.987 &    0.0 &    0.0 &  126.8 &    0.6 &    1.8 &    1.2 &   82 \\
& & & &    1.6 &    2.1 &   44.0 &    0.7 &    2.8 &    0.9 &   19 \\
 & \ref{FCOR_2} & 2004/05/24 & 4.993 &    0.0 &    0.0 &  183.8 &    1.3 &    2.2 &    1.9 &   49 \\
& & & &    2.7 &    2.8 &   23.3 &    1.1 &    2.3 &    1.4 &  155 \\
\object{J140051+521606} & \ref{FCOR_3} & 2000/12/26 & 1.408 &    0.0 &    0.0 &  170.0 &    0.5 &  182.3 &   95.9 &  102 \\
\object{J140942+360416} & \ref{FCOR_1} & 2001/05/28 & 1.659 &    0.0 &    0.0 &  104.1 &    0.2 &   10.7 &    4.9 &  159 \\
& & & &   -0.5 &   22.1 &   22.4 &    0.3 &   22.6 &    5.4 &   16 \\
 & \ref{FCOR_1} & 2004/06/05 & 1.665 &    0.0 &    0.0 &  111.1 &    0.3 &   13.0 &    5.8 &  169 \\
& & & &    1.4 &   26.9 &   10.3 &    0.2 &    6.1 &    4.4 &  148 \\
\object{J143521+505122} & \ref{FCOR_3} & 2000/12/26 & 1.408 &    0.0 &    0.0 &   75.2 &    0.6 &  161.0 &    9.3 &   69 \\
& & & &  128.8 &  -46.0 &   64.7 &    0.6 &  181.5 &   46.6 &  126 \\
\object{J150805+342323} & \ref{FCOR_3} & 2002/02/19 & 1.650 &    0.0 &    0.0 &    4.0 &    0.2 &    7.8 &    0.0 &   72 \\
& & & &  -97.5 & -144.4 &    6.1 &    0.4 &   42.2 &   30.5 &   47 \\
\object{J160246+524358} & \ref{FCOR_1} & 2001/05/27 & 1.659 &    0.0 &    0.0 &   26.0 &    0.1 &    0.0 &    0.0 &    0 \\
& & & &   13.8 &   -2.9 &  200.3 &    0.7 &   70.4 &   22.7 &  100 \\
& & & &  -73.0 &   16.2 &  222.0 &    0.8 &   64.0 &   32.8 &  106 \\
 & \ref{FCOR_1} & 2004/06/05 & 1.665 &    0.0 &    0.0 &   41.9 &    0.4 &    3.7 &    0.0 &  145 \\
& & & &   19.1 &   -7.8 &   20.5 &    0.6 &    6.1 &    2.8 &   49 \\
& & & &  -23.5 &    6.2 &    7.4 &    0.6 &    5.5 &    3.6 &   71 \\
& & & &  -36.2 &   -3.3 &   22.1 &    2.4 &   27.3 &   13.4 &   85 \\
& & & &  -63.6 &    8.8 &   26.8 &    1.5 &   15.9 &    8.3 &  132 \\
& & & & -104.6 &   43.2 &    2.2 &    0.6 &    6.3 &    3.5 &   11 \\
& & & &   65.7 &   -0.2 &    2.8 &    0.5 &    4.9 &    3.6 &   30 \\
 & \ref{FCOR_1} & 2004/05/24 & 4.993 &    0.0 &    0.0 &   37.7 &    0.1 &    4.1 &    3.0 &   75 \\
& & & &   19.1 &   -8.7 &   23.1 &    0.2 &    8.5 &    2.6 &  105 \\
& & & &  -17.5 &    8.9 &    5.1 &    0.2 &    7.6 &    2.9 &   74 \\
& & & &  -37.5 &    5.5 &    4.8 &    0.2 &   13.0 &    2.8 &   92 \\
\object{J161148+404020} & \ref{FCOR_3} & 2000/11/25 & 1.658 &    0.0 &    0.0 &  227.2 &    0.4 &  253.2 &  136.7 &   43 \\
& & & &  292.8 &  -15.9 &  214.3 &    0.5 &  304.5 &  222.0 &   22 \\
& & & & -591.8 &  -88.8 &   22.0 &    0.6 &  329.3 &  263.1 &  134 \\
\object{J171854+544148} & \ref{FCOR_1} & 2000/02/14 & 1.655 &    0.0 &    0.0 &   98.9 &    0.8 &    5.3 &    2.7 &    0 \\
& & & &   10.9 &  -50.0 &  151.4 &    1.8 &   19.2 &    5.5 &  169 \\
 & \ref{FCOR_1} & 2004/06/05 & 1.665 &    0.0 &    0.0 &  193.6 &    0.5 &    6.7 &    3.0 &    9 \\
& & & &    9.9 &  -45.2 &  167.9 &    0.7 &   17.3 &    2.9 &  180 \\
 & \ref{FCOR_1} & 2004/05/24 & 4.993 &    0.0 &    0.0 &   61.5 &    0.2 &    4.7 &    2.1 &  150 \\
& & & &   10.2 &  -47.4 &   26.0 &    0.2 &    5.0 &    3.2 &  115 \\
& & & &    9.9 &  -36.8 &   13.6 &    0.2 &    3.2 &    1.5 &  155 \\
& & & &    0.9 &  -19.8 &    3.8 &    0.3 &    9.1 &    0.0 &   98 \\
& & & &    0.4 &   20.0 &    2.3 &    0.2 &    5.1 &    0.0 &   90 \\
\object{J093609+331308} & \ref{FCOR_4} & 2000/03/02 & 4.987 &    0.0 &    0.0 &   51.7 &    0.3 &    1.5 &    0.3 &  142 \\
\object{J101636+563926} & \ref{FCOR_4} & 2000/11/09 & 1.658 &    0.0 &    0.0 &   57.2 &    0.2 &   64.4 &   26.5 &  154 \\
& & & &  136.9 & -196.6 &   39.2 &    0.1 &   66.5 &   10.7 &  137 \\
\object{J105731+405646} & \ref{FCOR_4} & 2000/03/02 & 4.987 &    0.0 &    0.0 &   14.8 &    0.1 &    0.5 &    0.2 &    1 \\
\object{J115727+431806} & \ref{FCOR_4} & 2001/01/02 & 1.408 &    0.0 &    0.0 &   43.2 &    0.2 &   55.5 &   44.2 &   76 \\
& & & & -185.3 &  604.5 &   59.4 &    0.2 &  215.0 &  131.0 &  165 \\
& & & &  134.8 &  188.5 &   30.4 &    0.2 &  100.8 &   55.2 &   10 \\
& & & &   41.4 &  423.5 &   58.4 &    0.3 &  232.8 &  187.6 &  155 \\
\object{J132513+395552} & \ref{FCOR_4} & 2000/03/02 & 4.987 &    0.0 &    0.0 &   16.2 &    0.1 &    0.4 &    0.0 &  176 \\
& & & &    2.2 &    5.4 &    3.0 &    0.2 &    1.0 &    0.8 &   86 \\
& & & &   -2.0 &   -4.2 &    5.6 &    0.7 &    6.1 &    2.0 &   15 \\
\object{J134035+444817} & \ref{FCOR_4} & 2000/03/02 & 4.987 &    0.0 &    0.0 &   78.3 &    0.2 &    0.2 &    0.2 &   13 \\
& & & &    0.7 &    3.2 &    5.7 &    0.3 &    1.8 &    0.7 &    9 \\
\object{J155927+533054} & \ref{FCOR_4} & 2000/12/18 & 1.408 &    0.0 &    0.0 &   83.0 &    0.8 &  936.4 &  390.6 &   33 \\
& & & &  560.2 & 1342.1 &   63.2 &    0.5 &  555.1 &  367.6 &    4 \\
\hline
\object{J071509+452555} & \ref{FCOR_5} & 2000/03/02 & 4.987 &    0.0 &    0.0 &   39.1 &    0.2 &    0.6 &    0.2 &  119 \\
& & & &    0.5 &   -1.2 &   11.3 &    0.2 &    0.8 &    0.1 &  162 \\
\object{J080454+433537} & \ref{FCOR_5} & 2000/02/14 & 1.655 &    0.0 &    0.0 &  194.1 &    0.4 &    6.7 &    3.7 &   70 \\
& & & &  -24.1 &   -1.8 &   71.3 &    0.4 &   15.5 &    3.2 &   63 \\
\object{J134158+541524} & \ref{FCOR_5} & 2002/02/19 & 1.650 &    0.0 &    0.0 &   66.3 &    0.8 &   67.9 &   40.9 &   50 \\
& & & & -270.9 &  467.1 &   10.9 &    0.4 &   33.8 &   16.3 &  108 \\
\end{longtable}


\begin{thebibliography}{}
\bibitem[2008]{Adelman-McCarthy08} Adelman-McCarthy, J.~K., et al.\ 2008, \apjs, 175, 297
\bibitem[1997]{Bicknell97} Bicknell, G.~V., Dopita, M.~A., \& O'Dea, C.~P.~O.\ 1997, \apj, 485, 112
\bibitem[2002]{Conway02} Conway, J.~E.\ 2002, New Astronomy Review, 46, 263
\bibitem[1990]{Fanti90} Fanti, R., Fanti, C., Schilizzi, R.~T., Spencer, R.~E., Nan Rendong, Parma, P., van Breugel, W.~J.~M., \& Venturi, T.\ 1990, \aap, 231, 333 
\bibitem[2003]{Fanti03} Fanti, C., \& Fanti, R.\ 2003, Radio Astronomy at the Fringe, ed. J.~A. Zensus, M.~H. Cohen, \& E. Ros, 300, 81
\bibitem[1999]{Fomalont99} Fomalont, E.~B.\ 1999, Synthesis Imaging in Radio Astronomy II, 180, 301 
\bibitem[2005]{Gugliucci05} Gugliucci, N.~E., Taylor, G.~B., Peck, A.~B., \& Giroletti, M.\ 2005, \apj, 622, 136 
\bibitem[1984]{Hodges84} Hodges, M.~W., Mutel, R.~L., \& Phillips, R.~B.\ 1984, \aj, 89, 1327
\bibitem[1974]{Jones74} Jones, T.~W., O'dell, S.~L., \& Stein, W.~A.\ 1974, \apj, 192, 261
\bibitem[2007]{Kaiser07} Kaiser, C.~R., \& Best, P.~N.\ 2007, \mnras, 381, 1548 
\bibitem[1981]{Kellermann81} Kellermann, K.~I., \& Pauliny-Toth, I.~I.~K.\ 1981, \araa, 19, 373
\bibitem[2008]{Komatsu08} Komatsu, E., et al.\ 2008, submitted to \apjs, ArXiv e-prints, 803, arXiv:0803.0547 
\bibitem[2007]{Liu07} Liu, X., Cui, L., Luo, W.-F., Shi, W.-Z., \& Song, H.-G.\ 2007, \aap, 470, 97
\bibitem[1991]{McMahon91} McMahon R.~G., Irwin M.~J., 1991, in Proceedings of the conference on `Digitised Optical Sky Surveys’, eds. H.~T. MacGillivray, E.~B. Thomson (Kluwer Acad. Publ., Dordrecht), p. 417
\bibitem[1985]{Mutel85} Mutel, R.~L., Hodges, M.~W., \& Phillips, R.~B.\ 1985, \apj, 290, 86
\bibitem[1996]{O'DeaBaum96} O'Dea, C.~P. \& Baum, S.~A.\ 1996, in GPS and CSS radio sources, ed. I.~A.~G. Snellen, R.~T. Schilizzi, H.~J.~A. R\"ottgering \& M.~N. Bremer, 241
\bibitem[1998]{O'Dea98} O'Dea, C.~P.\ 1998, \pasp, 110, 493
\bibitem[1998]{Owsianik98} Owsianik, I., \& Conway, J.~E.\ 1998, \aap, 337, 69 
\bibitem[1998]{Owsianik98b} Owsianik, I., Conway, J.~E., \& Polatidis, A.~G.\ 1998, \aap, 336, L37 
\bibitem[1982]{Peacock82} Peacock, J.~A., \& Wall, J.~V.\ 1982, \mnras, 198, 843
\bibitem[2003]{Polatidis03} Polatidis, A.~G., \& Conway, J.~E.\ 2003, Publications of the Astronomical Society of Australia, 20, 69
\bibitem[1996]{Readhead96} Readhead, A.~C.~S., Taylor, G.~B., Xu, W., Pearson, T.~J., Wilkinson, P.~N., \& Polatidis, A.~G.\ 1996, \apj, 460, 612
\bibitem[2000a]{Snellen00a} Snellen, I.~A.~G., Schilizzi, R.~T., \& van Langevelde, H.~J.\ 2000, \mnras, 319, 429
\bibitem[2000b]{Snellen00b} Snellen, I.~A.~G., Schilizzi, R.~T., Miley, G.~K., de Bruyn, A.~G., Bremer, M.~N., R\"ottgering, H.~J.~A.\ 2000, \mnras, 319, 445 
\bibitem[2002]{Snellen02} Snellen, I.~A.~G., Lehnert, M.~D., Bremer, M.~N., \& Schilizzi, R.~T.\ 2002, \mnras, 337, 981
\bibitem[2003]{Snellen03} Snellen, I.~A.~G., Mack, K.-H., Schilizzi, R.~T., \& Tschager, W.\ 2003, Publications of the Astronomical Society of Australia, 20, 38 
\bibitem[2004]{CORALZ} Snellen, I.~A.~G., Mack, K.-H., Schilizzi, R.~T., \& Tschager, W.\ 2004, \mnras, 348, 227 
\bibitem[2007]{Spergel} Spergel, D.~N., et al.\ 2007, \apjs, 170, 377 
\bibitem[1997]{Stanghellini97} Stanghellini, C., O'Dea, C.~P., Baum, S.~A., Dallacasa, D., Fanti, R., \& Fanti, C.\ 1997, \aap, 325, 943
\bibitem[1998]{Stanghellini98} Stanghellini, C., O'Dea, C.~P., Dallacasa, D., Baum, S.~A., Fanti, R., \& Fanti, C.\ 1998, \aaps, 131, 303
\bibitem[2002]{Stanghellini02} Stanghellini, C., Liu, X., Dallacasa, D., \& Bondi, M.\ 2002, New Astronomy Review, 46, 287
\bibitem[2000]{Taylor00} Taylor, G.~B., Marr, J.~M., Pearson, T.~J., \& Readhead, A.~C.~S.\ 2000, \apj, 541, 112
\bibitem[2002]{Tingay02} Tingay, S.~J., et al.\ 2002, \apjs, 141, 311 
\bibitem[2000]{Tschager00} Tschager, W., Schilizzi, R.~T., R{\"o}ttgering, H.~J.~A., Snellen, I.~A.~G., \& Miley, G.~K.\ 2000, \aap, 360, 887
\bibitem[1998]{Tzioumis98} Tzioumis, A.~K., et al.\ 1998, IAU Colloq.~164: Radio Emission from Galactic and Extragalactic Compact Sources, 144, 179
\bibitem[1997]{White97} White, R.~L., Becker, R.~H., Helfand, D.~J., \& Gregg, M.~D.\ 1997, \apj, 475, 479
\bibitem[1994]{Wilkinson94} Wilkinson, P.~N., Polatidis, A.~G., Readhead, A.~C.~S., Xu, W., \& Pearson, T.~J.\ 1994, \apjl, 432, L87
\end{thebibliography}
\end{document}